\documentclass[twocolumn]{aastex701}

\usepackage{multirow} \usepackage{graphicx}	
\usepackage{amsmath}	

\usepackage{savesym}  
\savesymbol{splitbox}
\usepackage[export]{adjustbox}
\restoresymbol{TXF}{splitbox}
\newcommand{\fracA}{{f_{\rm{AGN}}}}
\newcommand\tb{\textcolor{black}}
\usepackage{booktabs}
\usepackage{multirow}

\usepackage{booktabs}
\usepackage{CJK}

\begin{document}
\begin{CJK*}{UTF8}{gbsn}

\title{MEOW: The increase in the obscured AGN fraction in mid-infrared from $0\leq z\leq 6$ with JWST MIRI}

\author[0000-0001-8174-6389]{Teodora-Elena Bulichi}
\affiliation{Massachusetts Institute of Technology, Kavli Institute for Astrophysics and Space Research, 77 Massachusetts Avenue, Cambridge, MA 02139, USA}
\affiliation{Department of Physics, Massachusetts Institute of Technology, 77 Massachusetts Avenue, Cambridge, MA 02139, USA}
\email[show]{teob1823@mit.edu}  

\author[0000-0002-9393-6507]{Gene C. K. Leung} 
\affiliation{Massachusetts Institute of Technology, Kavli Institute for Astrophysics and Space Research, 77 Massachusetts Avenue, Cambridge, MA 02139, USA}
\email{fakeemail4@google.com}

\author[0000-0003-2895-6218]{Anna-Christina Eilers}
\affiliation{Massachusetts Institute of Technology, Kavli Institute for Astrophysics and Space Research, 77 Massachusetts Avenue, Cambridge, MA 02139, USA}
\affiliation{Department of Physics, Massachusetts Institute of Technology, 77 Massachusetts Avenue, Cambridge, MA 02139, USA}
\email{eilers@mit.edu}

\author[0000-0003-4528-5639]{Pablo G. P\'erez-Gonz\'alez}
\affiliation{Centro de Astrobiolog\'{\i}a (CAB), CSIC-INTA, Ctra. de Ajalvir km 4, Torrej\'on de Ardoz, E-28850, Madrid, Spain}
\email{fakeemail4@google.com}

\author[0000-0001-6813-875X]{Guillermo Barro}
\affiliation{Department of Physics, University of the Pacific, Stockton, CA 90340 USA}
\email{fakeemail4@google.com}

\author[0000-0001-8519-1130]{Steven L. Finkelstein}
\affiliation{Department of Astronomy, The University of Texas at Austin, Austin, TX, USA}
\affiliation{Cosmic Frontier Center, The University of Texas at Austin, Austin, TX, USA}
\email{fakeemail4@google.com}

\author[0000-0002-9921-9218]{Micaela B. Bagley}
\affiliation{Department of Astronomy, The University of Texas at Austin, Austin, TX, USA}
\affiliation{Cosmic Frontier Center, The University of Texas at Austin, Austin, TX, USA}
\email{fakeemail4@google.com}

\author[0000-0002-6610-2048]{Anton M. Koekemoer}
\affiliation{Space Telescope Science Institute, 3700 San Martin Drive, Baltimore, MD 21218, USA}
\email{fakeemail4@google.com}

\author[0000-0001-8534-7502]{Bren E. Backhaus}
\affiliation{Department of Physics and Astronomy, University of Kansas, Lawrence, KS 66045, USA}
\email{fakeemail4@google.com}

\author[0000-0001-5414-5131]{Mark Dickinson}\affiliation{NSF's National Optical-Infrared Astronomy Research Laboratory, 950 N. Cherry Ave., Tucson, AZ 85719, USA}
\email{fakeemail4@google.com}

\author[0000-0001-7201-5066]{Seiji Fujimoto}
\affiliation{David A. Dunlap Department of Astronomy and Astrophysics, \\ University of Toronto, 50 St. George Street, Toronto, Ontario, M5S 3H4, Canada}
\affiliation{Dunlap Institute for Astronomy and Astrophysics, 50 St. George Street, Toronto, Ontario, M5S 3H4, Canada}
\email{fakeemail4@google.com}

\author[0000-0001-9440-8872]{Norman A. Grogin}
\affiliation{Space Telescope Science Institute, 3700 San Martin Drive, Baltimore, MD 21218, USA}
\email{nagrogin@stsci.edu}

\author[0000-0002-8360-3880]{Dale D. Kocevski}
\affiliation{Department of Physics and Astronomy, Colby College, Waterville, ME 04901, USA}
\email{fakeemail4@google.com}

\author[0000-0003-1581-7825]{Ray A. Lucas}
\affiliation{Space Telescope Science Institute, 3700 San Martin Drive, Baltimore, MD 21218, USA}
\email{fakeemail4@google.com}

\author[0000-0001-9879-7780]{Fabio Pacucci}
\affiliation{Center for Astrophysics $\vert$ Harvard \& Smithsonian, Cambridge, MA 02138, USA}
\email{fabio.pacucci@cfa.harvard.edu}

\author[0000-0003-3382-5941]{Nor Pirzkal}
\affiliation{Space Telescope Science Institute, 3700 San Martin Drive, Baltimore, MD 21218, USA}
\email[]{}

\author[0000-0002-9712-0038]{Elia Pizzati}
\affiliation{Center for Astrophysics $\vert$ Harvard \& Smithsonian, Cambridge, MA 02138, USA}
\email{fakeemail4@google.com}

\author[0000-0002-4544-8242]{Jan-Torge Schindler}
\affiliation{Hamburger Sternwarte, University of Hamburg, Gojenbergsweg 112, D-21029 Hamburg, Germany}
\email{fakeemail4@google.com}

\author[0000-0003-1006-924X]{Alberto Traina}
\affiliation{Istituto Nazionale di Astrofisica (INAF) - Osservatorio di Astrofisica e Scienza dello Spazio (OAS), via Gobetti 101, I-40129 Bologna, Italy}
\email{fakeemail4@google.com}

\author[0000-0001-8835-7722]{Guang Yang (杨光)}
\affiliation{Nanjing Institute of Astronomical Optics \& Technology, Chinese Academy of Sciences, Nanjing 210042, China}
\email{fakeemail4@google.com}

\begin{abstract}

Obscured active galactic nuclei (AGN) are often invoked to explain the rapid emergence of young quasars at high redshift and are crucial for building a complete census of AGN activity and black hole growth. The advent of the \textit{James Webb Space Telescope} (\textit{JWST}) extends the discovery space for obscured AGN into the mid-infrared (mid-IR) with unprecedented precision through reprocessed dust emission. In this work, we use deep \textit{JWST} Mid-Infrared Instrument (MIRI) imaging from the MIRI Early Obscured AGN Wide Survey (MEOW), together with existing \textit{JWST} Near Infrared Camera (NIRCam), spectroscopic, and \textit{Hubble Space Telescope} imaging data, to identify a previously unrecognized population of obscured AGN out to $z \approx 6$. Using spectral energy distribution (SED) modeling of the MIRI-detected sources, we identify 883 AGN over an area of $\approx131\,\mathrm{arcmin^2}$ and construct the AGN bolometric luminosity function, including both obscured and unobscured sources, across five redshift bins. We find an excess in AGN abundance relative to UV-selected AGN luminosity functions, indicating a substantial obscured population missed by optical/UV surveys, with the inferred obscured fraction increasing with redshift and reaching $\approx 98$--$99\%$ in our highest-redshift bin, $4.5 < z < 6$. We also find higher AGN abundances and obscured fractions than X-ray-based studies, consistent with a previously unrecognized population of heavily obscured, Compton-thick AGN revealed by mid-IR selection. These results suggest that a large fraction of supermassive black hole growth at early times occurs during heavily obscured phases largely inaccessible at other wavelengths.

\end{abstract}



\section{Introduction}

The existence of quasars powered by supermassive black holes (SMBHs) with masses attaining $\sim10^{9} \, \mathrm{M}_\odot$ within the first billion years of cosmic history (e.g., \citealt{Mortlock2011, Mazzucchelli2017, FanBanadosSimcoe2023, Eilers2020}) presents a significant challenge for theoretical models of black hole growth. In the standard picture, and following the Soltan argument \citep{Soltan1982}, SMBHs are thought to assemble their mass through gas accretion, observed during phases of active galactic nucleus (AGN) or quasar activity. Even under the idealized assumption of continuous accretion at the Eddington limit, the time required to grow these objects from stellar remnants is comparable to the age of the Universe at these redshifts, $t_Q \sim 10^8-10^9\,\mathrm{yr}$, assuming a radiative efficiency corresponding to the cosmic average $\epsilon_\mathrm{rad} \approx 10\%$ (e.g., \citealt{Volonteri2010,Inayoshi2020,Eilers2021, Pacucci2022}). Thus, how such massive black holes grew so rapidly, and under what physical conditions this growth occurred, remain key open questions.

The SMBH growth timescale at high redshift can be constrained by measuring the ionizing imprint signatures of quasars on the surrounding intergalactic medium (IGM), since the gas responds to changes in the ionizing output over a finite timescale. In practice, this has been probed using Ly$\alpha$ proximity zones ($z \approx 6$, \citealt{Eilers2017}),  Ly$\alpha$ damping wings at higher redshifts ($z \gtrsim 6$, \citealt{Davies2019}), and, more recently, spatially extended ionized nebulae around quasars \citep{Durovcikova2025}. All of these tracers consistently point toward surprisingly short lifetimes, as low as $t_Q \lesssim 10^4\,\mathrm{yr}$ \citep{Eilers2017,Eilers2018,Davies2019,Eilers2020,Andika2020,Morey2021,Satyavolu2023,Durovcikova2024,Durovcikova2025}. Moreover, population-level constraints on UV-bright duty cycles have independently supported these unexpectedly short timescales \citep{Arita2023, Eilers2024, Pizzati2024, Schindler2025, Huang2026}. Taken together, these findings underscore the pressing need to reassess our understanding of SMBH growth over cosmic history.

This intriguing tension has motivated two main scenarios that could explain the observations and the early SMBH growth, namely (i) radiatively inefficient ``super-Eddington'' accretion, allowing for a more rapid mass build-up( e.g., \citealt{Inayoshi2016,Pezzulli2016,Begelman2017,Regan2019,Davies2020,Husko2024,Obuchi2025}), possibly even starting from heavier seeds resulted from e.g., direct collapse (e.g., \citealt{Haiman1997,BrommLoeb2003,Begelman2006,Volonteri2008,Wise2019,Lupi2021,Pacucci2026}) and/or (ii) heavily dust-obscured phases that attenuate the ionizing output and suppress UV-bright signatures of accretion (e.g., \citealt{Hopkins2008,Polletta2008,Merloni2014,Vito2018,Lyu2024}). In the latter case, recent duty cycle estimates imply that $\sim 100$ obscured quasars must exist for every unobscured one at $z \gtrsim 6$ in order to reconcile the observed SMBH masses with the observed growth timescales \citep{Eilers2024,Schindler2025,Huang2026}, far exceeding the obscured fractions of $\sim 20-50\%$ inferred in the local Universe, which rely predominantly on hard X-ray observations \citep{Burlon2011,Iwasawa2012,Ueda2014,Buchner2015,Liu2017,Zappacosta2018,Iwasawa2020}. 

While theoretical predictions \citep{Trebitsch2019,Ni2020,Vito2022,Bennett2024}, indirect measurements from the interstellar medium (ISM) constraints \citep{Gilli2022}, \tb{and decades of multi-wavelength searches for obscured AGN \citep[see reviews by][]{Hickox2018,Sajina2022}} point towards a redshift evolution of the obscured fraction (see also \citealt{Peca2025}), constraining its value at high redshift (i.e., $z \gtrsim 3$) has remained challenging: strong gas and dust columns, coupled with cosmological dimming, render X-ray surveys increasingly incomplete at early times, with detection efficiencies dropping to $\lesssim 30$--$40\%$ for obscured AGN \citep{Vito2018}, while also making redshift constraints more challenging due to their faint or absent optical/UV counterparts. Moreover, recent radio studies suggest that the population of Compton-thick AGN at early epochs ($z \approx 3$) may be substantially underestimated in Cosmic X-ray Background models, pointing to a heavily obscured AGN population that is larger by a factor of $\approx2-3$ than currently accounted for \citep{Mazzolari2026}.

An alternative tracer of obscured accretion emerges in the rest frame near-infrared, redshifted to mid-infrared (mid-IR) at high redshift, where the hot circumnuclear dust re-radiates the absorbed AGN emission (e.g., \citealt{Lacy2004,Stern2005,Alonso-Herrero2006A,Donley2012,Assef2013,Ling2026}). However, before the \textit{James Webb Space Telescope (JWST)}, mid-IR studies at $z \gtrsim 2$ were severely limited by the restricted wavelength coverage and sensitivity of previous facilities, particularly beyond $\sim8\,\mu\mathrm{m}$. This limitation has been overcome by \textit{JWST}'s Mid-Infrared Instrument (MIRI; \citealt{Wright2023}), which offers nine broadband filters in the wavelength range $5$--$26\,\mu\mathrm{m}$, covering the hot-dust spectral features of AGN out to $z \approx 8$. Relative to its predecessor, \textit{Spitzer}, MIRI achieves $\sim 10$--$100\times$ higher sensitivity and $\sim 8\times$ finer angular resolution at comparable wavelengths, enabling the detection and characterization of obscured SMBH growth with unprecedented precision (e.g., \citealt{Yang2023,Lyu2024,Hsieh2025}).

\tb{Moreover, pre-\textit{JWST} discoveries of individual high-redshift obscured or reddened AGN candidates have further highlighted the importance of this population. For example, \cite{Fujimoto2022} identified GNz7q, a compact dust-enshrouded source at $z\simeq7.19$ that may represent a transition phase between dusty starbursts and unobscured quasars, while \cite{Endsley2023} confirmed COS-87259, a heavily obscured, hyperluminous AGN candidate at $z\simeq6.83$. These findings suggest that obscured accretion may already be common in the reionization era, but a homogeneous statistical census remains necessary to quantify its prevalence, now possible with \textit{JWST}.}

In this paper, we present a homogeneous measurement of the obscured AGN fraction out to $z \approx 6$ using the MIRI Early Obscured-AGN Wide Survey (MEOW, \#5407, PIs Leung, Endsley, Finkelstein; \citealt{Leung2026}). This program adds deep MIRI 10 and 21 $\mu$m imaging over $\sim 95\,\mathrm{arcmin}^2$ in the Great Observatories Origins Deep Survey (GOODS; \citealt{Giavalisco}) fields. We  supplement our analysis with data from SMILES, which covers $\approx 34 \,\mathrm{arcmin}^2$  in GOODS-S \citep{Rieke2024}, with a total area of $\approx 65\,\mathrm{arcmin}^2$ split equally between GOODS-N and GOODS-S. Leveraging MIRI's sensitivity to hot-dust emission, in combination with extensive \textit{JWST} NIRCam imaging, and NIRSpec spectroscopy from JADES \citep{Einsenstein2026}, NIRCam WFSS from FRESCO \citep{Oesch2023}, as well as archival coverage from the Hubble Space Telescope (HST), we robustly identify a previously unrecognized population of obscured AGN at early cosmic times. These observations provide new constraints on SMBH growth pathways that could alleviate the short quasar lifetime tension (e.g., \citealt{Eilers2018, Eilers2020,Schindler2025,Durovcikova2025}).

\tb{The overall framework of the analysis presented in this paper is illustrated in Figure~\ref{fig:roadmap}.} The \tb{text is organized} as follows: \S~\ref{Sec:data} discusses the data acquisition and processing, \S~\ref{Sec:AGN_sample} describes the AGN identification through SED fitting, used simultaneously for redshift determinations and completeness corrections; \S~\ref{Sec:LF} presents the methodology for the luminosity function measurements, which we use in \S~\ref{Sec:results} to compute the obscured AGN fraction; 
in \S~\ref{Sec:discussion}, we place our results in the context of previous work and discuss the implications for early SMBH growth. Finally, \S~\ref{Sec:conclusions} summarizes our main findings.

Throughout the paper, we adopt the flat $\Lambda$CDM \cite{Planck2016} cosmology:  $H_0 = 67.66\,\rm km\,s^{-1}\,Mpc^{-1}$, $\Omega_M = 0.3097$, and $\Omega_\Lambda = 0.6889$.

\section{Data}\label{Sec:data}

In this section, we describe the data products used for the AGN sample identification (\S~\ref{Sec:AGN_sample}). A detailed description of the \textit{JWST} MEOW program, data reduction, source detection and photometry are available in \cite{Leung2026}, and here we provide a summary of the key procedures.

\subsection{MIRI observations}\label{sec:MEOW}

In this study, we make use of deep MIRI imaging, in F1000W, and F2100W, within the GOODS-N and GOODS-S fields, through the MEOW \textit{JWST} program.
MEOW is designed to uncover the population of heavily obscured supermassive black holes that remain elusive at optical/UV wavelengths. By coupling these new MIRI observations with existing NIRCam slitless spectroscopy, deep multi-wavelength imaging in the GOODS fields, and archival \textit{HST} data, the survey identifies AGN through their characteristic hot-dust emission, a well-established tracer of obscured accretion at early cosmic times (see \S~\ref{Sec:AGN_sample}).

MEOW comprises 74~hours of MIRI imaging over an area of $95~\mathrm{arcmin^2}$, using the F1000W and F2100W filters. Each pointing includes exposure times of approximately $720$~s in F1000W and $3000$~s in F2100W, employing the FASTR1 readout and a four-point dither pattern. 

Data in GOODS-S are supplemented by F1000W and F2100W observations from SMILES (JWST \#1207, PI: George Rieke, \citealt{Rieke2024}). SMILES comprises 15 separate pointings over a total area of $\approx 34\,\mathrm{arcmin^2}$ in GOODS-S, and similarly uses the FASTR1 readout together with a four-point dither pattern. The total exposure time of $\approx 2.2\,\mathrm{h}$ is divided among eight MIRI filters (F560W, F770W, F1000W, F1280W, F1500W, F1800W, F2100W, and F2550W). For consistency with MEOW, we make use only of the F1000W and F2100W imaging, with integration times of $642$ and $2184$ seconds, respectively. Additionally, we employ the same source detection methodology as in MEOW on the two SMILES F1000W and F2100W images (see \S~\ref{sect:sources_ident}). 

We additionally incorporate MIRI imaging data in the GOODS fields publicly available in September 2025. These include F1000W observations from PIDs 2926 and 4762 in GOODS-N, and PIDs 1283, 4498, 6511 in GOODS-S. These observations fall completely within the MEOW or SMILES footprints. Lastly, the total survey areas in GOODS-N and GOODS-S are approximately equal, $\approx 65\,\mathrm{arcmin^2}$.

\begin{figure*}
    \includegraphics[width=0.95\textwidth]{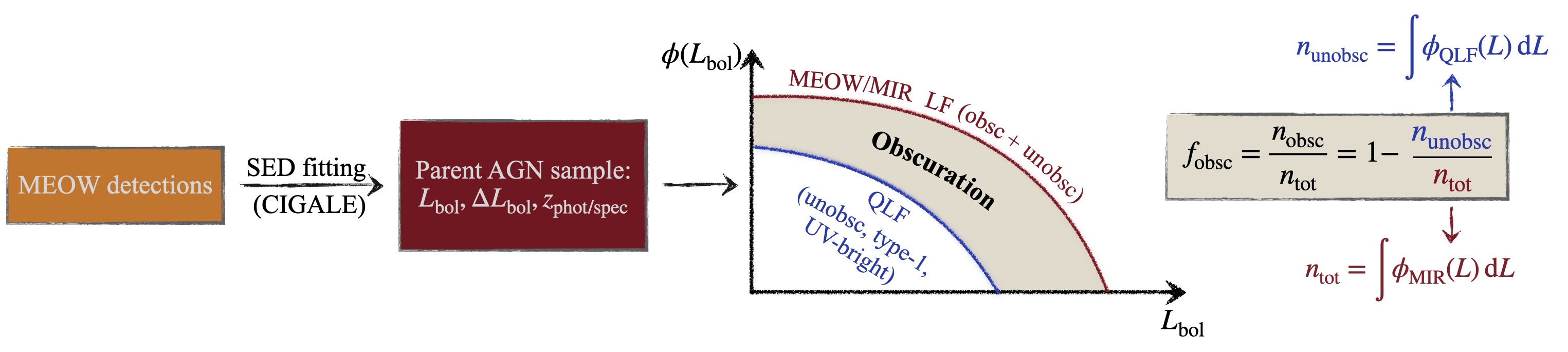}
    \centering
    \caption{\textbf{Framework for quantifying the AGN obscured fraction.} We illustrate the layout in characterizing obscuration, from the MEOW source detections (\S~\ref{sect:sources_ident}), followed by the identification of AGN, and corresponding bolometric luminosities and uncertainties $L_\mathrm{bol},\Delta L_\mathrm{bol}$ (\S~\ref{sect:AGN_identification}), as well as photometric redshift determinations (\S~\ref{sec:z-det}) through SED fitting. These measurements enable the construction of AGN luminosity functions across five redshift bins, incorporating both obscured and unobscured populations (\S~\ref{Sec:LF}). Finally, we quantify the obscured fraction as a function of luminosity and redshift by evaluating the excess of MEOW AGN relative to type-1, unobscured quasars (\S~\ref{sect:f_obsc}). }
    \label{fig:roadmap}

\end{figure*} 

\subsection{MIRI data reduction}\label{sec:data_reduction}
The data reduction was carried out with the JWST Rainbow pipeline, following the techniques described in Appendix A of \cite{Perez-Gonzalez2024}. More specific details about the MEOW and the supplemental SMILES dataset are given in \cite{Leung2026}. In short, the data reduction is based on the official \textit{JWST} pipeline (v1.19.41, jwst\_1413.pmap), with custom modifications in the background subtraction. Specifically, we introduce additional steps, in the ``super-background'' subtraction described in the following paragraph, to improve the treatment of spatially varying and small-scale background structures, particularly at the shortest MIRI wavelengths.

We first processed the data through the standard \textit{JWST} pipeline to obtain an initial mosaic, which we used to detect sources with \textsc{SExtractor} \citep{1996A&AS..117..393B} and generate a segmentation map. From this, we built source masks that flag all detected objects and their extended emission. These source masks were then mapped back to the native pixel grid of each individual calibrated frame, ensuring that only background regions remained unmasked. With astrophysical sources removed, we modeled and subtracted large-scale background structure and then constructed a super-background image by median-combining the masked, background-normalized exposures. This super-background was scaled and subtracted from each frame, significantly improving background flatness and suppressing low-frequency instrumental patterns (see also \citealt{Yang2023,Alberts2024,Perez-Gonzalez2024b,Backhaus2025,Ostlin2025}). \tb{After the data reduction, the resulting 1$\sigma$ depths are 0.0998 $\mu \rm{Jy}$ in F1000W, and 0.718 $\mu \rm{Jy}$ in F2100W.}

\subsection{MIRI photometry}\label{sect:sources_ident}

We detected the MIRI sources using \textsc{SExtractor} version 2.28.2, individually on each of the two filters (F1000W and F2100W), \tb{and constructed the MIRI parent sample by requiring sources to be detected in both MIRI filters, with positional matches within 0.5 arcseconds (see \S~\ref{sect:nircam_hst}, \S~\ref{sect:parent_sample})}. We employed a hot-cold detection scheme, following the framework described in \cite{Galametz2013}, \tb{adopting \texttt{DEBLEND\_MINCONT} values of 0.001 and 0.02 for the hot detection scheme, for F1000W and F2100W respectively, and 0.2 for the cold detection scheme}. The cold configuration was optimized for bright, extended sources using conservative detection thresholds, minimum-area requirements, and deblending parameters, while the hot configuration employed more aggressive settings to recover faint sources. The final catalog was constructed by retaining all sources from the cold detection and including hot-detected sources only if their positions lay outside a dilated cold segmentation map. Spurious detections significantly narrower than the point-spread function were removed based on the fraction of flux enclosed within a small aperture. 

The MIRI photometry, for both MEOW and the supplemental SMILES F1000W and F2100W images, was measured in circular apertures with radii of $0.3''$ in F1000W and $0.5''$ in F2100W, and aperture corrections were applied using the MIRI instrument team flux calibration products \citep{Gordon2025}, based on the reference file \texttt{jwst\_miri\_apcorr\_0014.fits}. The photometric uncertainties were estimated by placing the aperture of the same size in non-overlapping empty/sky regions of the mosaic. Specifically, we computed the median absolute deviation of all fluxes and converted it to an equivalent Gaussian uncertainty by multiplying by 1.48. We then applied the same aperture correction as for the fluxes. Lastly, to account for different depths in different parts of the mosaic, we multiplied the error of each source by the ratio between the value of the error map at the source location and the median value of the error map. 

\subsection{NIRCam and HST photometry} \label{sect:nircam_hst}

The NIRCam and HST photometry are based on the UNICORN project catalogs (v0.89 for GOODS-S, v0.9 for GOODS-N; Finkelstein et al., in prep.), which includes all HST/ACS and NIRCam data in the MEOW footprints. This is inclusive of the FRESCO survey data in the F182M, F210M and F444W filters, and HST data in ACS WFC F435W, F606W, F775W, F814W, F850L; WFC3/IR F105W, F125W, F140W, F160W from the GOODS and CANDELS surveys, all of which cover the full MEOW footprint.  Imaging from JADES is available in parts of the MEOW footprint, covering F090W, F115W, F150W, F200W, F210M, F277W, F335M, F356W, F410M, and additionally F210M and F480M in GOODS-S. 

The UNICORN catalog uses \textsc{SExtractor} and the hot-cold detection scheme, on a combined F277W+F356W used as the primary detection image, supplemented by F444W as the secondary detection image, as parts of the MEOW footprint are not covered by F277W+F356W. The NIRCam detections were matched with the MIRI sources (\S~\ref{sect:sources_ident}), requiring a detection in each F444W, F1000W and F2100W for our final parent sample (see also \S~\ref{sect:parent_sample}). The photometric procedures follow the methodology presented in \cite{Finkelstein2024}, with updates presented in \cite{Taylor2025}, and we refer the reader to these papers for more details. 

For each source, we have at least a total of 11 filters to constrain the SED, and the HST coverage aids in identifying the Lyman break, and thus helps constraining the redshift for high-$z$ objects more accurately (see \S~\ref{sec:z-det}).


\subsection{Redshift determination} \label{sec:z-det}

\begin{figure}
    \includegraphics[width=0.475\textwidth]{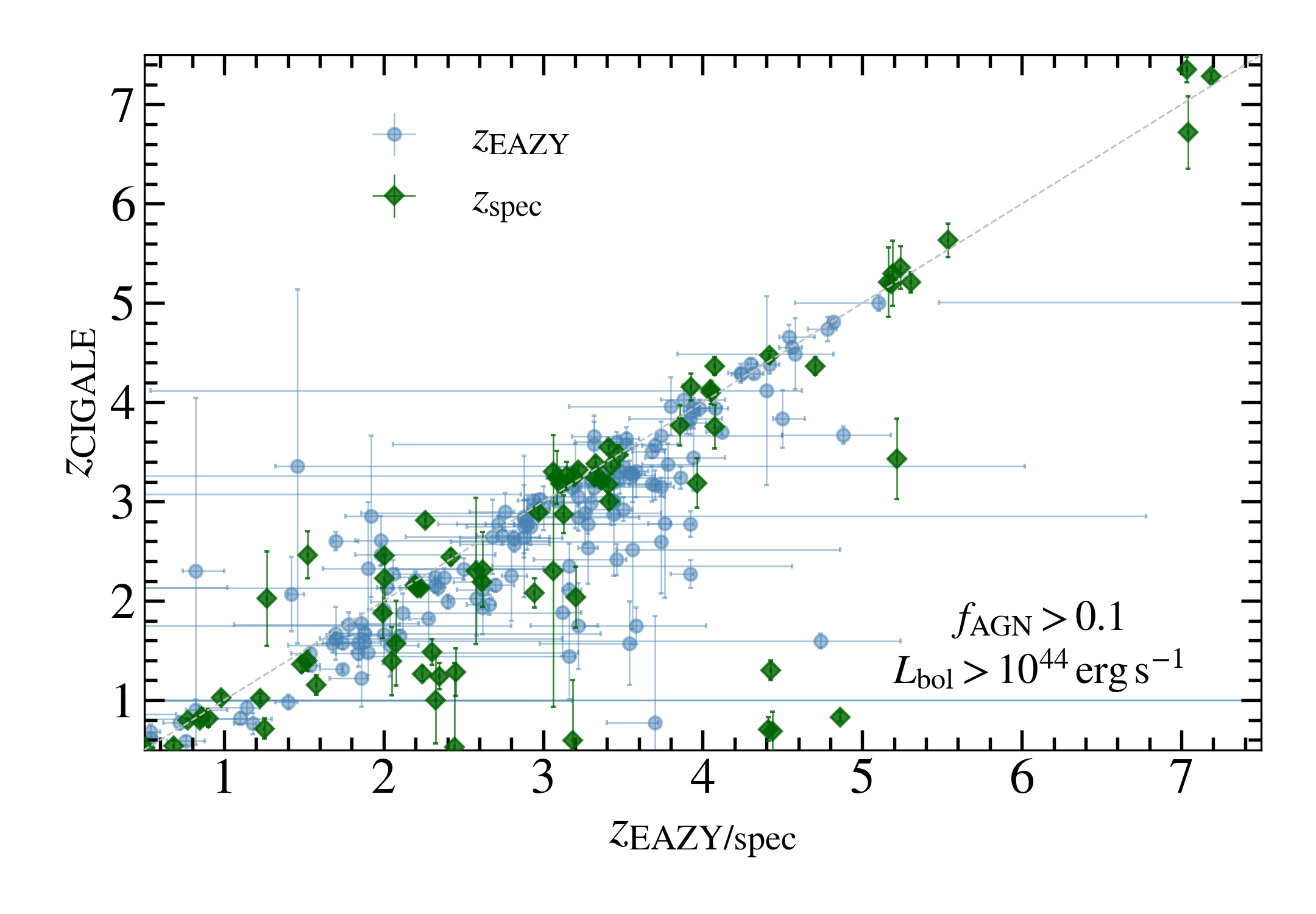}
    \caption{\textbf{\texttt{CIGALE} redshift estimates compared to \texttt{eazy} (blue) and spectroscopic measurements (green).} We show only the sources included in the analysis presented in this paper (\S~\ref{Sec:LF}), namely those with $f_\mathrm{AGN} > 0.1$ and $L_\mathrm{bol} > 10^{44}\,\mathrm{erg\,s^{-1}}$. Overall, the \texttt{CIGALE} redshifts are in good agreement with both the \texttt{eazy} and spectroscopic measurements; however, we identify 41 outliers, as well as a bias toward lower redshift estimates from \texttt{CIGALE} (see text). Of the 254 sources, 83 have secure spectroscopic redshifts, which are adopted for the analysis in this paper. For the remaining 171, we rely on the \texttt{CIGALE} redshift estimates, noting that the choice between \texttt{CIGALE} and \texttt{eazy} does not affect the population-level results of this work (Appendix~\ref{sec:app_LF}). }
    \label{fig:zphot-spec}

\end{figure} 

While for most sources ($67\%$ of the AGN sample, see \S~\ref{sect:AGN_identification}) we have to rely on photometric redshifts, we use the spectroscopic redshift wherever available from CANDELS \citep{Grogin2011,Koekemoer2011,Kodra2023}, JADES \citep{Einsenstein2026} and FRESCO \citep{Oesch2023}. All three programs provide coverage in both GOODS-S and GOODS-N, across multiple redshifts ($z \gtrsim 4.5$ for FRESCO, extended to lower redshifts in CANDELS and JADES). We identify counterparts to our sample by requiring an angular separation smaller than 0.2 arcseconds. For all the other sources, we obtain two photometric redshifts for each source, one using the code \texttt{eazy} on the NIRCam and HST photometry (\S~\ref{sect:nircam_hst}), part of the UNICORN catalogs, which fits linear combinations of galaxy templates to the observed photometry to minimize $\chi^2$ across a redshift grid \citep{eazy}; and the second directly from \texttt{CIGALE} as part of the SED fitting (\S~\ref{sec:CIGALE-setup}). \tb{We emphasize, however, that the main analysis presented in this paper relies exclusively on the \texttt{CIGALE} photometric redshifts, when spectroscopic redshifts are not available. This choice is motivated by the fact that \texttt{CIGALE} is better suited to the goals of this study, as it models the AGN contribution self-consistently, whereas \texttt{eazy} does not include AGN templates. The \texttt{eazy} counterparts are presented only to provide a comparison between two independent photometric-redshift estimates and we further explore the impact of this choice on our results in Appendix~\ref{sec:app_LF}.}

We show the results from these two methods in Fig.~\ref{fig:zphot-spec}, highlighting only the sources with $\fracA > 0.1$ (see \S~\ref{sect:AGN_identification}), and AGN bolometric luminosity $L_\mathrm{bol} > 10^{44}\,\mathrm{erg\,s^{-1}}$, used subsequently in our analysis (\S~\ref{Sec:LF} and \S~\ref{Sec:results}). 
In Figure~\ref{fig:zphot-spec}, where available, we show the spectroscopic redshift in lieu of \texttt{eazy}, and also show for comparison the corresponding \texttt{CIGALE} redshift determinations. We note that for all the analysis of this paper, we re-ran the SED fitting with the spectroscopic redshifts fixed, where available, and we only show the corresponding CIGALE photo-$z$ in Fig.~\ref{fig:zphot-spec} to illustrate the accuracy of redshift determinations as part of the SED fitting. 

First, we report an overall good agreement between the photometric redshift estimates from \texttt{CIGALE} and the corresponding spectroscopic values, with $\approx 88\%$ of the sources within $|\Delta z| < 1$ (i.e., the size of the luminosity function redshift bins, \S~\ref{Sec:LF}), and a median of $|\Delta z|_\mathrm{med} \approx 0.17$ from the spectroscopic counterpart. We find that if \texttt{CIGALE} misidentifies the redshift, it tends to generally underestimate it, and we identify seven clear outliers where the code yields photometric redshifts of $z_\mathrm{CIGALE} \lesssim 1$ for sources with spectroscopic redshifts $z_\mathrm{spec} \gtrsim 3$. Upon closer inspection, three of these seven sources exhibit large uncertainties in the HST photometry blueward of the Balmer break. The remaining four show degeneracy between low- and high-redshift solutions, with \texttt{CIGALE} preferentially assigning higher weight to the low-redshift fit. The same is true for the four sources with $z_\mathrm{CIGALE}\approx 1.5$ and $z_\mathrm{spec}\approx 2.2$. \tb{Despite these discrepancies, we find that their impact on our LF measurements is negligible. Incorporating photometric redshift uncertainties directly into the LF computations, by perturbing each redshift according to its corresponding uncertainty, leaves the resulting luminosity functions essentially unchanged (\S~\ref{sec:LF_non-par}). The small number of clear outliers therefore has no significant effect at the population level. Furthermore, adopting spectroscopic redshifts when available helps minimize the uncertainties propagated into the subsequent analyses.}

A comparison of the photometric redshifts from \texttt{CIGALE} and \texttt{eazy} shows overall good agreement, though we identify 30 clear outliers out of 171 sources. In six cases, the \texttt{eazy} redshift probability distribution exhibits a secondary peak consistent with the \texttt{CIGALE} solution. An additional 11 \texttt{eazy} estimates are poorly constrained, with uncertainties spanning $\Delta z_\mathrm{eazy} \gtrsim 3$. For the remaining 13 sources, the \texttt{CIGALE} fits are well constrained, suggesting that the discrepancy likely arises from template degeneracies in \texttt{eazy}, although more systematic effects may also contribute, including the tendency for \texttt{CIGALE} to favor lower-redshift solutions.

\tb{As mentioned before, }we adopt the \texttt{CIGALE} redshifts for sources without spectroscopic measurements. This choice is motivated by the code's ability to derive redshifts consistently with the full SED fitting, \tb{accounting self consistently for the AGN and galaxy host spectral features, as well as dust re-processing in IR}. We emphasize, however, that our results are not sensitive to the choice between \texttt{eazy} and \texttt{CIGALE}, as the redshift discrepancy for most sources is less than the width of the redshift bins used subsequently in the luminosity function calculations (\S~\ref{Sec:LF}), and the small number of outliers has no statistically significant effect on the overall population. A detailed comparison is presented in Appendix~\ref{sec:app_LF}.

\section{AGN sample}\label{Sec:AGN_sample}

In this section, we discuss the steps undertaken to obtain the complete MIRI AGN sample, which represents the basis of the subsequent analysis of this paper (\S~\ref{Sec:LF}, \S~\ref{Sec:results}). We first describe the parent sample on which we performed SED fitting using \texttt{CIGALE} \citep{Boquien2019,Yang2020}, then outline the AGN selection criteria, and conclude by assessing the AGN sample completeness.

\subsection{Parent sample}\label{sect:parent_sample}

We use the MIRI detections (\S~\ref{sect:sources_ident}) to identify and characterize the parent AGN sample (\S~\ref{Sec:AGN_sample}), with the aim of constructing the redshift and luminosity evolution of the AGN obscured fraction, as depicted in Fig.~\ref{fig:roadmap}. 

First, in order to obtain sufficient photometric coverage for SED fitting (see \S~\ref{sec:CIGALE-setup}), we match the MIRI detections with the corresponding NIRCam detections (see also \S~\ref{sect:nircam_hst}), \tb{requiring a matching radius/separation of less than 0.5 arcseconds. For the final parent catalog, we require sources to be detected in each of the three filters: F444W, F1000W, and F2100W.} We note that this selection \tb{would} remove any potential NIRCam dark sources, \tb{but upon closer inspection, we did not find any NIRCam dark objects in our sample}. We identify a total of 1518 sources in GOODS-N, and 1191 sources in GOODS-S. We note that the difference between the two fields is consistent with the results reported in \cite{Backhaus2025}, where they find a lower source count in SMILES (GOODS-S), compared to MEGA/CEERS(EGS). Nonetheless, the overall distribution of sources in type (see \S~\ref{sect:AGN_identification}) and redshift bins is consistent between the two fields (see Fig.~\ref{fig:N_hist}), with the exception of $z \approx 5$, where GOODS-N is over-dense \citep{Daddi2009,Casey2016,Miller2018,Lin2025}. Lastly, we find that the field-to-field variations do not impact our luminosity functions (\S~\ref{Sec:LF}, \tb{see also \citealt{Vidal2026}}) and/or obscured fractions (\S~\ref{sect:f_obsc}), including in the highest redshift bin where small-number statistics lead to larger uncertainties.

\subsection{SED fitting} \label{sec:CIGALE-setup}

\begin{table*}
\centering
\caption{\texttt{CIGALE} model parameters}
\label{tab:cig}
\begin{tabular}{llll} \hline\hline
Module & Parameter & Symbol & Values \\
\hline
\multirow{2}{*}{\shortstack[l]{Star formation history\\
                               \texttt{sfhdelayed} }}
    & Stellar e-folding time & $\tau_{\rm star}$ & 0.5, 1, 2, 3, 4, 5 Gyr\\
    & Stellar age & $t_{\rm star}$  
            & 0.1, 0.2, 0.5, 1, 2, 3, 4, 5 Gyr\\ 
\hline
\multirow{2}{*}{\shortstack[l]{Simple stellar population\\ 
    \texttt{bc03} }}
    & Initial mass function & $-$ & \cite{Chabrier2003} \\
    & Metallicity & $Z$ & 0.02 \\
\hline
\multirow{2}{*}{\shortstack[l]{Nebular emission\\
                               \texttt{nebular} }} 
    & Ionization parameter & $\log U$ & $-2.0$\\
    & Gas metallicity & $Z_{\rm gas}$ & 0.02 \\
\hline
\multirow{3}{*}{\shortstack[l]{Dust attenuation \\ 
                \texttt{dustatt\_modified\_starburst} }}
    & \multirow{2}{*}{\shortstack[l]{Color excess of nebular lines}} & \multirow{2}{*}{\shortstack[l]{$E(B-V)_{\rm line}$}} & \multirow{2}{*}{\shortstack[l]{0, 0.02, 0.05, 0.1, 0.2,\\
                              0.3, 0.4, 0.5, 0.7, 0.9, 1.0}} \\\\
    & ratio between line and continuum $E(B-V)$ & $\frac{E(B-V)_{\rm line}}{E(B-V)_{\rm cont}}$ & 1 \\
\hline
\multirow{4}{*}{\shortstack[l]{Galactic dust emission \\ \texttt{dl2014} }}
    & PAH mass fraction & $q_{\rm PAH}$ & 0.47, 2.5, 7.32 \\
    & Minimum radiation field & $U_{\rm min}$ & 0.1, 1.0, 10, 50 \\
    & \multirow{2}{*}{\shortstack[l]{Fraction of PDR emission}} & \multirow{2}{*}{\shortstack[l]{$\gamma$}} & \multirow{2}{*}{\shortstack[l]{0.01, 0.02, 0.05, \\
                                            0.1, 0.2, 0.5, 0.9}} \\\\
\hline
\multirow{5}{*}{\shortstack[l]{AGN (UV-to-IR) emission \\ \texttt{skirtor2016} }}
    & Average edge-on optical depth at $9.7 \mu$m & $\tau_{9.7}$ & 3, 5, 7, 9, 11 \\
    & Viewing angle & $\theta_{\rm AGN}$ & 70$^\circ$ (type2) \\
    & \multirow{2}{*}{\shortstack[l]{AGN contribution to \tb{the 3--30 $\mu\rm{m}$} luminosity}} & \multirow{2}{*}{\shortstack[l]{$\fracA$}} & \multirow{2}{*}{\shortstack[l]{0, 0.05,  
    0.1--0.9 (step 0.1), 0.99  }}\\\\
    & Wavelength range where $\fracA$ is defined & $\lambda_{\rm AGN}$ & 3--30 $\mu$m \\
    &  $E(B-V)$ for the extinction in the polar direction & $EBV$ & 0, 0.05, 0.1, 0.15, 0.2, 0.3 \\
\hline
    \multirow{2}{*}{\shortstack[l]{Redshift$+$IGM \\ \texttt{redshifting} }} 
    &  \multirow{2}{*}{\shortstack[l]{ Source redshift }} & \multirow{2}{*}{\shortstack[l]{ $z$ }} & \multirow{2}{*}{\shortstack[l]{0--8.0} (step 0.1)} \\\\
\hline
\end{tabular}
\begin{flushleft}
{\sc Note.} --- For parameters not listed here, default values are being adopted.
\end{flushleft}
\end{table*}

\begin{figure*}
    \includegraphics[width=0.95\textwidth]{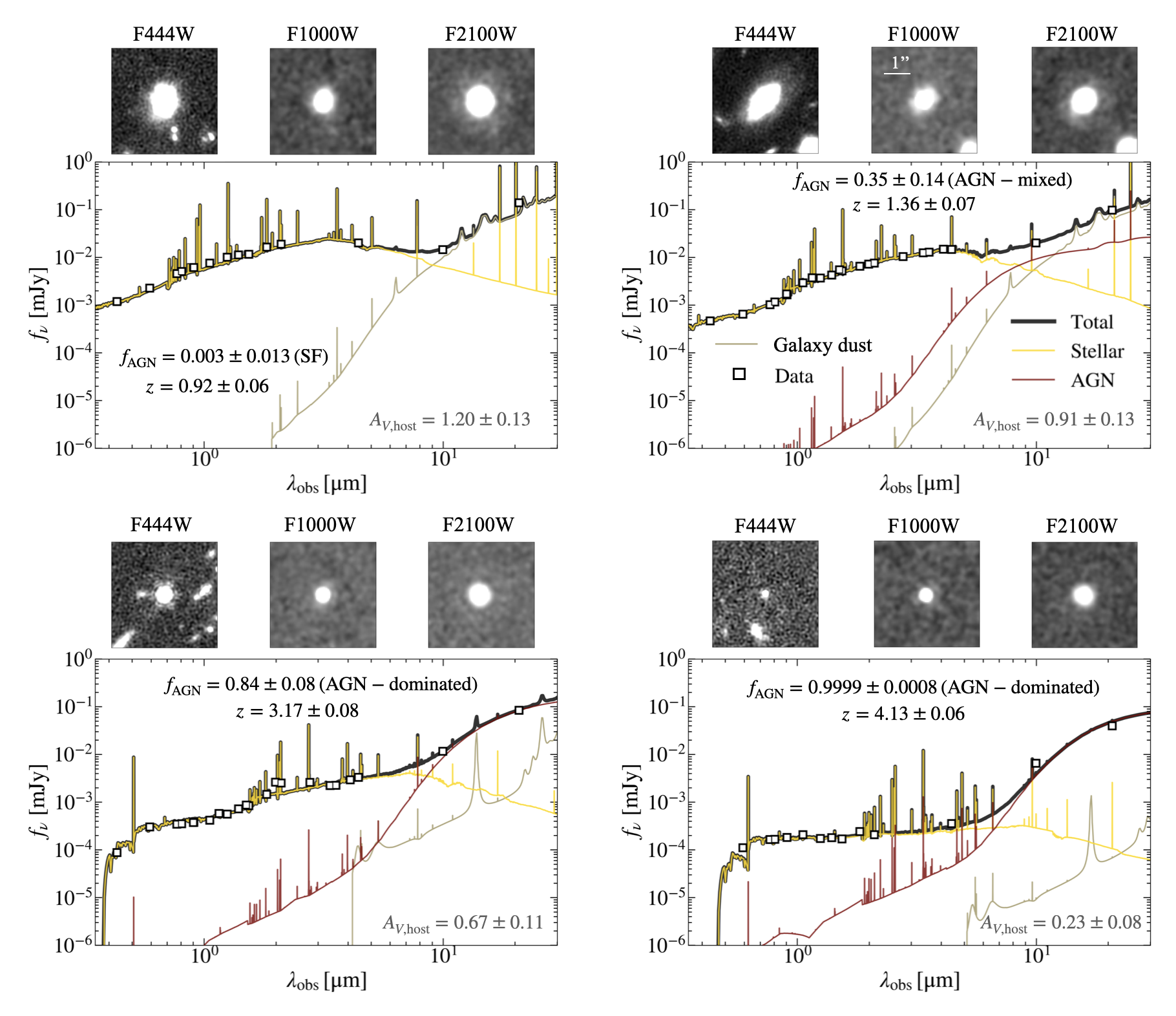}
    \centering
    \caption{\textbf{Examples of SED fits, and corresponding cut-outs in the F444W, F1000W, F2100W filters, for a SF object at $z = 0.92$ (top-left), and three AGN, with increasing $f_\mathrm{AGN}$ (quoted on the plot) at $z = 1.36$ (top-right), $z = 3.17$ (bottom-left) and $z=4.13$ (bottom-right).} Photometric measurements are plotted as open black squares; uncertainties are included but not visible due to the high signal to noise. In the first two cases, the negligible AGN contribution in the star-forming system and the modest enhancement above the dust component in the top-right object support their classifications. The last two objects are clearly AGN-dominated, highlighted by the mid-IR bump that can only be explained through the hot AGN circumnuclear dust, which exceeds the galactic dust responsible for attenuating the blue stellar continuum. }
    \label{fig:cut_outs}

\end{figure*}

In order to identify the AGN, as well as their key properties (e.g., luminosity, obscured fraction), we perform SED modeling on all the sources, using \texttt{CIGALE} v2022.1 \citep{Boquien2019,Yang2020,Yang2022}, on the MIRI and existing NIRCam/HST filters (see \S~\ref{sect:sources_ident}). We show all the customized \texttt{CIGALE} parameters in Table~\ref{tab:cig}, following a similar approach to \cite{Yang2023}, and provide a short description of each component below. All the other parameters not listed in the table adopt the default values of \texttt{CIGALE}. \tb{In total, \texttt{CIGALE} fits 13 free parameters, with an input of at least 11 photometric points, and maximum 20, depending on the overlap with existing NIRCam/HST data (see \S~\ref{sect:nircam_hst}).}

For the stellar population modeling, we employ a standard delayed star formation history, with the $e$-folding time allowed to vary between 0.5-5 Gyr, and the stellar age between 0.1-5 Gyr. We adopt a \cite{Chabrier2003} initial mass function (IMF), and a constant metallicity $Z = 0.02$. We additionally include a nebular emission module \citep{Villa-Velez2021}, which assumes the same gas metallicity as the stellar metallicity, $Z = 0.02$, and a constant ionization parameter $\rm{log} U = -2$ for the \ion{H}{2} regions.

The effects of galactic dust are modeled assuming a \cite{Calzetti2000} dust attenuation law, extended into the far-UV (i.e., at wavelengths shorter than the nominal Calzetti range, down to the Lyman break) following the prescriptions of \cite{Leitherer2002}; as well as considering galactic dust emission according to the model described in \cite{Draine2014}. The latter is assumed to equal the attenuated luminosity of starlight, and is split between diffuse emission and emission from photodissociation regions (PDRs), corresponding to star formation sites. In both regions, the mass fraction of polycyclic aromatic hydrocarbon (PAH) is set by the same parameter $q_\mathrm{PAH}$, which we allow to vary between three values: 0.47, 2.5 and 7.32. The radiation fields of the two dust emission regions are controlled by the minimum radiation parameter ($U_{\mathrm{min}}$), and the $\gamma$ parameter (see Table~\ref{tab:cig}). Namely, the dust radiation from the diffuse region is set by $U_{\rm{min}}$, while the PDR's dust emission ranges from $U_{\rm{min}}$ to a fixed maximum value $U_{\rm{max}} = 10^7$; and the relative contribution of the two components is controlled by the parameter $\gamma$.

The AGN emission, covering the UV to IR wavelength range, is described via \texttt{CIGALE}'s \texttt{skirtor2016} module, assuming a clumpy torus model described in \cite{Stalevski2012,Stalevski2016}. We allow the optical depth at 9.7 $\mu\rm{m}$ to vary between all available values: 3, 5, 7, 9, 11. For all the fits, we assume that the AGN are type-2, set by the viewing angle (see Table~\ref{tab:cig}). We emphasize, however, that this choice is independent of how obscuration is quantified later (\S~\ref{sect:f_obsc}; see also Fig.~\ref{fig:roadmap}). Its only impact is on the bolometric luminosity estimate, through the different AGN contribution at bluer wavelengths in type-1 and type-2 templates, but the differences are mild as the luminosity is strongly constrained by the rest frame near-IR signature. As explained in previous studies \citep{Ni2021,Yang2021,Zou2022,Yang2023}, including type-1 models would cause a strong degeneracy between AGN and stellar contributions in the UV-optical regime, making it more difficult to identify the obscured AGN, which are the focus of this work. We \tb{also} expect the number of type-1 AGN to be rarer, especially at high redshift (e.g., \citealt{Lacy2015}). We note, however, that this choice does not affect our population results, such as luminosity functions (see Appendix~\ref{sec:app_LF}). 

The relative AGN contribution to the \tb{3--30 $\mu\rm{m}$} luminosity is set by $\fracA$, which covers the AGN dust emission, \tb{similar to the methodology of \cite{Yang2023}}. \tb{We emphasize that our customized wavelength range differs from the default \texttt{CIGALE} definition of $\fracA$, which is computed over 8--1000 $\mu\rm{m}$, and that alternative definitions have also been adopted in the literature. For example, \citealt{Kirkpatrick2015} compute an analogous $\fracA$ over 5--15 $\mu\rm{m}$. Despite these differences in wavelength coverage, we find good agreement for the sources used in our luminosity functions that overlap with \cite{Kirkpatrick2015}. Of the 10 overlapping AGN, nine are assigned the same classification, either mixed or AGN-dominated, in both studies. The only exception has $\fracA \approx 0.3$ in our sample, corresponding to a mixed classification, but $\fracA = 0.07$ in \cite{Kirkpatrick2015}, corresponding to a star-forming classification. We note that this source is faint, with an inferred luminosity of order $\sim 10^{44}\,\mathrm{erg\,s^{-1}}$ in both studies, and therefore its inferred AGN contribution may be particularly sensitive to the wavelength coverage and sensitivity of the observations.} Similar to \cite{Yang2023}, we model both the dust on the torus scale, as well as the polar dust extinction via the $EBV$ parameter.

Lastly, we allow \texttt{CIGALE} to fit the redshift of all sources (see \S~\ref{sec:z-det}), using a linear grid $z = 0-8$, in steps of $\Delta z = 0.1$. For the sources with spectroscopic redshift, we fix those values for all the SED fitting.

We show four examples of SED fitting in Figure~\ref{fig:cut_outs}, along with the image cut-outs in the two MEOW filters, F1000W and F2100W, as well as the corresponding F444W NIRCam image for each source. The reference sources were chosen at four different redshifts $z = 0.92$, $z = 1.36$, $z = 3.17$ and $z = 4.13$, as well as with different AGN contributions, quantified by $\fracA$. In all cases, we highlight the individual contributions of the stellar component, the AGN, and the galaxy dust to the total SED. While both the galaxy dust and AGN components emit in the mid-IR, which can lead to degeneracies in the fitting and potentially affect the AGN identification, the use of energy balance in \texttt{CIGALE} provides an additional constraint. In particular, the galaxy dust emission is tied to the attenuation of the blue stellar continuum \tb{(which we quantify via $A_{V,\mathrm{host}}$, shown in Fig.~\ref{fig:cut_outs})}, helping to distinguish it from AGN-heated dust. This is especially evident in the bottom-right panel of Fig.~\ref{fig:cut_outs}, where the mid-IR bump can only be explained by the AGN dust component: the stellar continuum shows little attenuation at bluer wavelengths \tb{($A_{V,\mathrm{host}} = 0.23 \pm 0.08)$}, indicating very low levels of host dust. Therefore, we expect any mid-IR degeneracy to have only a minor impact at the population level, affecting mostly the sources with \tb{subdominant} AGN contributions \tb{(i.e., $\fracA \lesssim 0.5$)}. \tb{Additionally, we verified that including sub-mm constraints for the high-redshift sources does not affect the SED fits or the resulting $\fracA$ classifications. This suggests that our AGN identifications are not driven by confusion with heavily obscured star-forming systems. This point is discussed further in \cite{Leung2026}.}

\tb{Lastly, we note that the four sources presented in the 5 $\times$ 5'' cut-outs above the SED fits exhibit diverse morphologies in the NIRCam imaging, while the corresponding MIRI emission appears more compact for the sources with higher inferred AGN contributions. However, given the small number of examples shown here, these cut-outs are not sufficient to draw firm conclusions about the broader source population, and we defer a detailed population-level analysis of AGN morphology to future work.}

Based on the \texttt{CIGALE} SED fitting results, we further identify and characterize the parent AGN sample in the following subsection. We additionally visually inspected all sources and removed spurious objects (about $\approx 1\%$ of the total objects), such as saturated stars, also confirmed by the high $\chi^2$ reported in \texttt{CIGALE}.

\subsection{AGN identification}\label{sect:AGN_identification}

\begin{figure}
    \includegraphics[width=0.475\textwidth]{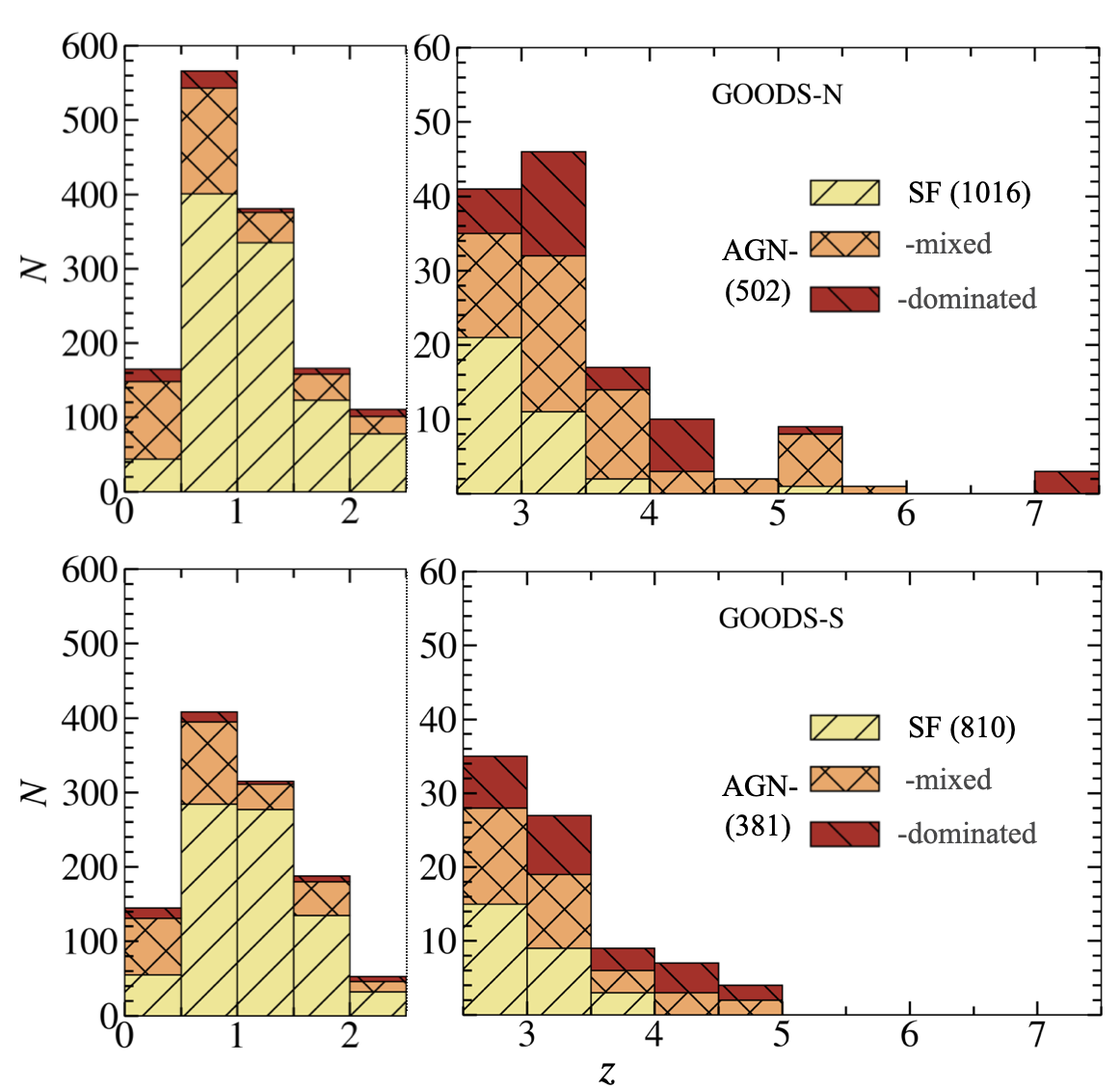}
    \centering
    \caption{\textbf{Stacked histogram of the redshift distribution for sources in GOODS-N (top) and GOODS-S (bottom).} Star-forming systems dominate both fields, except at $z \gtrsim 4$, where the MIRI selection becomes increasingly sensitive to AGN and mixed sources. Despite the difference in overall number counts (see text), the relative fractions of star-forming, composite, and AGN sources remain broadly consistent across the two fields. }
    \label{fig:N_hist}
\end{figure} 

\begin{figure}
    \includegraphics[width=0.5\textwidth]{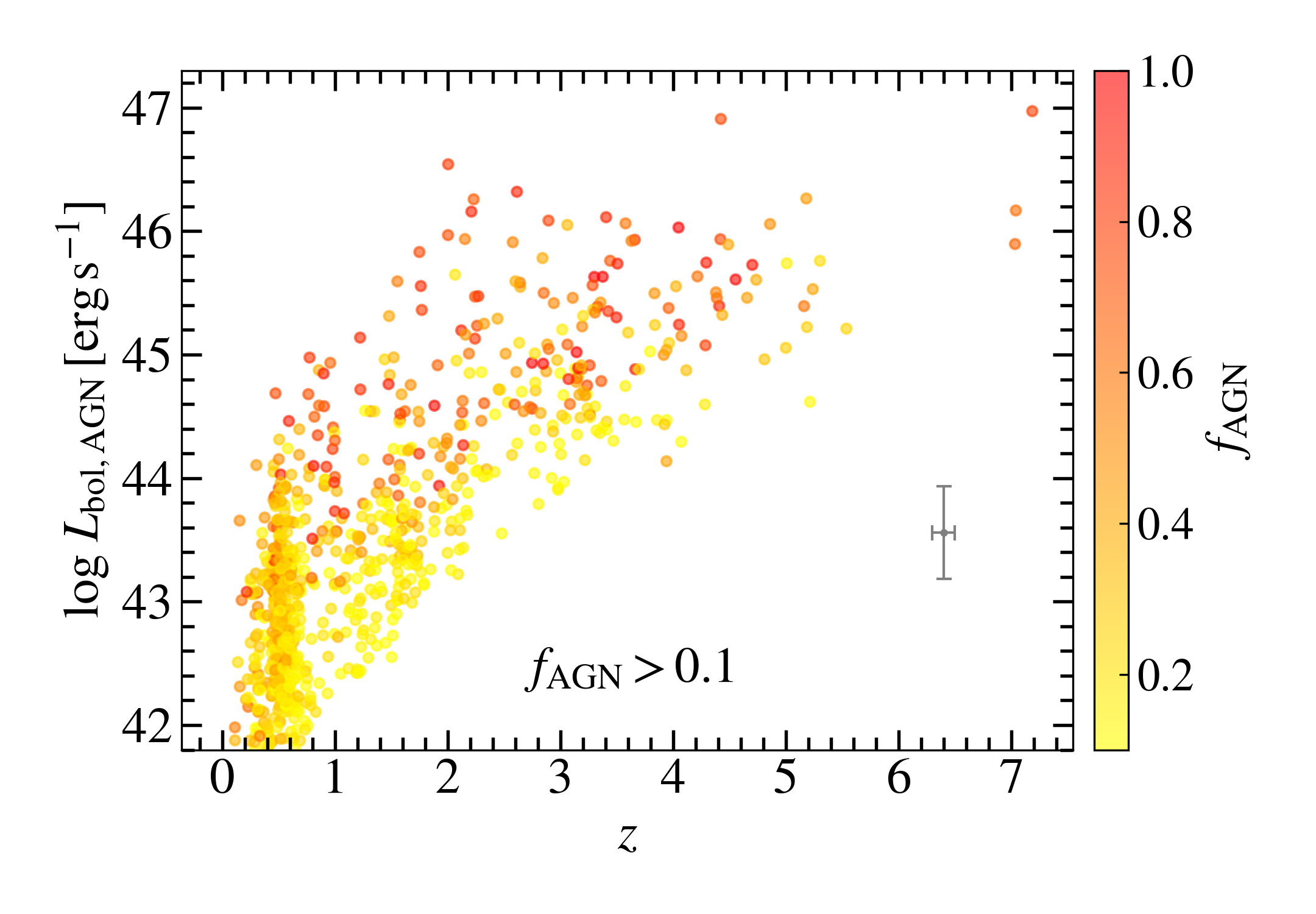}
    \caption{\textbf{Bolometric \tb{AGN} luminosity distribution across redshift, color-coded by $f_\mathrm{AGN}$.}
    We show only sources with $f_\mathrm{AGN} > 0.1$ (mixed and AGN-dominated), and include typical error bars in grey, corresponding to $|\Delta z| \approx 0.1$ and $|\Delta \mathrm{log} L_\mathrm{bol}| \approx 0.4$ dex. Overall, the MEOW depth allows us to detect down to $L_\mathrm{bol} \approx 10^{44} \,\mathrm{erg\,s^{-1}}$ across cosmic time, and for the identification of high redshift AGN up to $z\lesssim 7$, with 18 AGN at $z \gtrsim 4.5$. }
    \label{fig:Lbol_z}
\end{figure}

In order to identify the AGN in our sample, we use the Bayes \texttt{CIGALE} output $\fracA$ (see \S~\ref{sec:CIGALE-setup}). Similar to \cite{Yang2023} (see also \citealt{Durodola2025,Ling2026}), we separate our sources between star-forming (SF): $\fracA < 0.1$; AGN-mixed: $0.1 \le \fracA < 0.5$; AGN-dominated: $\fracA \ge 0.5$. We identify 1017 SF sources, and 502 AGN (dominated + mixed) in GOODS-N; and 810 SF and 381 AGN in GOODS-S. We show the distribution of different source types over redshift in Figure~\ref{fig:N_hist}.

The relative distribution of source types and redshifts is consistent between the two fields, with roughly two thirds of the population classified as star-forming, and one third AGN (-mixed and -dominated). The latter will be further used in the AGN luminosity function (\S~\ref{Sec:LF}) calculations and subsequent analysis for quantifying obscuration (\S~\ref{Sec:results}, see also Fig.~\ref{fig:roadmap}). We note that most of the SF sources are present at $z \lesssim 4$, and beyond that the MIRI selection effects preferentially identify mixed and AGN-dominated sources, due to the pronounced hot-dust signatures. This leads to a first statistical sample of 15 high-redshift AGN (\tb{$4.5 \lesssim z \lesssim 6$}) that show strong obscuration effects (\S~\ref{Sec:results}, detailed further in \citealt{Leung2026}). 

\tb{Although our AGN identification is based on the mid-IR signatures of hot circumnuclear dust, the selection is not intrinsically tied to a particular viewing angle. Provided that the system hosts a detectable hot-dust component, our method should recover the AGN irrespective of orientation, and is therefore sensitive to both obscured and unobscured systems. To verify this explicitly for unobscured sources, we cross-matched our AGN sample with the broad-line AGN identified by JADES in GOODS-N and GOODS-S \citep{Juodvzbalis2026}, as well as with the optical broad-line AGN in GOODS-S reported by \cite{Szokoly2004}. Compared to the JADES type-1 AGN sample, we recover three of the four sources that lie above our detection limits and within the MEOW footprint: one in GOODS-S and two in GOODS-N. The only exception is GN-73488 from \cite{Juodvzbalis2026}. Upon closer inspection, this source is classified as a little red dot (LRD) in \cite{Kocevski2025} (UNICORN ID 18850), and is absent from our parent sample because it is not detected in MIRI, consistent with the weak or absent mid-IR emission often reported in LRDs \citep{Williams2024,Perez-Gonzalez2024,Leung2024} In addition, we recover all five broad-line AGN from \cite{Szokoly2004} that overlap with our footprint. This comparison confirms that our selection does recover known type-1 AGN when they satisfy the same mid-IR detection criteria applied to the full sample.}

Figure~\ref{fig:Lbol_z} shows the redshift distribution of bolometric luminosity, of the AGN component, for each source with $\fracA > 0.1$. We adopt as bolometric luminosity the \texttt{CIGALE} output \texttt{bayes.agn.accretion\_power}, which traces only the intrinsic, unextincted AGN accretion-disk emission, averaged over all viewing directions, and explicitly excludes any stellar or dust emission from the host galaxy. In \texttt{CIGALE}, this quantity is computed as the likelihood-weighted Bayesian mean over the marginalized probability distribution function built from the full grid of fitted SED models. We color-code the points by $\fracA$, finding that luminous AGN show generally high $\fracA$, and vice-versa (see also \citealt{Hamblin2025,Ling2026}).

\tb{We note that the three AGN at $z \gtrsim 7$ are not included in the luminosity function calculation, as the extremely small number of sources in this regime would make the resulting space-density estimates highly uncertain and sensitive to individual objects. We therefore conservatively restrict the highest-redshift luminosity function bin to $4.5 \lesssim z \lesssim 6$, where the sample provides a more statistically meaningful constraint. For the same reason, we do not interpret the apparent gap between $z \approx 5.7 \rightarrow 7$ as evidence for an intrinsic absence of AGN, but rather as a consequence of the limited survey volume and low expected source counts at these redshifts.}

A table with the AGN properties of our entire sample is presented in Appendix~\ref{sec:tab_data}, available in machine readable form in the online version of the article.  

\subsection{Sample completeness}\label{sec:completeness}

We assess the completeness of our sample in two components: the photometric detection completeness and the SED-based identification completeness. For the photometric detection completeness, we use source injection simulations. Specifically, we inject sources within a range of F1000W and F2100W fluxes, matched from the AGN templates, and recover them using the same source detection procedures to calculate the detection completeness in each filter as a function of fluxes. The final detection completeness is the product of the completeness in the two filters at the corresponding fluxes/bolometric luminosities.

We additionally model SED completeness, in order to account for sources that are not recovered as AGN-dominated/-mixed when performing SED fitting. Namely, we define five different redshift bins, used later for the luminosity function (\S~\ref{Sec:LF}): $0.5 < z < 1.5$, $1.5 < z < 2.5$, $2.5 < z < 3.5$, $3.5 < z < 4.5$ and  $4.5 < z < 6$. We first construct representative AGN-dominated and mixed source SEDs. For each redshift–luminosity bin, we compute the median SED, separately for AGN-dominated and AGN-mixed populations, normalized at a rest-frame wavelength of 3 $\mu \rm{m}$ to mitigate biases driven by intrinsically bright sources. We then scale the median SEDs to a given bolometric luminosity by leveraging the empirical relation between $L_{3\mu\mathrm{m}}$ and $L_\mathrm{bol}$, obtained from a linear fit across the full sample.

Using these normalized templates, we generate mock spectra across a grid in redshift and $L_\mathrm{bol}$ and extract synthetic photometry in the observed filters. To realistically reflect measurement uncertainties, each flux point is perturbed five times using the median photometric error in that filter. Each perturbed mock catalogue is then processed through the full \texttt{CIGALE} pipeline. The SED completeness in each redshift luminosity bin is defined as the fraction of injected AGN or mixed sources that are recovered as $f_\mathrm{AGN} > 0.1$ by our selection criteria, relative to the total number of simulated sources. The final completeness is the product of the detection completeness and SED completeness, and the completeness curves are available in \cite{Leung2026}.

Finally, we quantify the effective volume for both mixed and AGN sources, using the following relation:

\begin{equation*}
V(L_\mathrm{bol})_\mathrm{eff} = \int \frac{dV}{dz} \, C(L_\mathrm{bol}, z) \, dz,
\end{equation*}

\noindent where $C(L_\mathrm{bol}, z)$ is the combined completeness from source injection and SED completeness, and the final effective volume is the weighted average of the effective volume for mixed and AGN sources in that bin, respectively:

\begin{equation*}
\tb{V(L_\mathrm{bol})_\mathrm{final}} = \frac{N_\mathrm{mixed}V_\mathrm{eff,mixed}+N_\mathrm{AGN}V_\mathrm{eff,AGN}}{N_\mathrm{mixed}+N_\mathrm{AGN}},
\end{equation*}

\noindent where $N_\mathrm{mixed}$ and $N_\mathrm{AGN}$ represent the numbers of mixed sources and AGN sources, respectively.

\tb{We report the completeness values in the redshift and luminosity bins used for the LF calculations (\S~\ref{Sec:LF}) in Table~\ref{tab:data}. The parameter space probed in this study is generally highly complete ($\gtrsim93\%$), with the exception of the lowest-luminosity bin at $4.5 < z < 6$, where the completeness decreases to 21\% for mixed sources and 11\% for AGN-dominated sources; further details are presented in \cite{Leung2026}. Although these points therefore require substantial completeness corrections in the LF calculation, our main conclusions do not rely on this luminosity range, as it is not covered by the comparison quasar luminosity function of \cite{Kulkarni2019}. Similarly, the second luminosity bin at $4.5 < z < 6$ is only moderately complete, with completeness below 80\%, but it is not used in the main obscured-fraction calculations presented in \S~\ref{Sec:results}.}


\section{Bolometric luminosity function} \label{Sec:LF}

In this section, we discuss the methods employed to infer the AGN luminosity function, in five redshift bins, using both a non-parametric approach, as well as a double power law fit. Subsequently, we will use our luminosity functions in comparison with the existing type-1 UV-bright luminosity functions, to quantify obscuration (\S~\ref{Sec:results}, see also Fig.~\ref{fig:roadmap}).


\subsection{Non-parametric approach}\label{sec:LF_non-par}
Given our final, complete AGN sample (\S~\ref{sec:completeness}), with associated bolometric luminosities and corresponding uncertainties from the SED fitting (\S~\ref{sec:CIGALE-setup}), we can compute the bolometric luminosity function (LF) across redshift. We calculate the LFs in five redshift bins, $z = 0.5 - 6$, using the \texttt{CIGALE} redshift estimates, or spectroscopic redshifts where available (\S~\ref{sec:z-det}).

First, we calculate the non-parametric luminosity function following the methodology of \cite{Finkelstein2015, Finkelstein2023, Leung2023}. Specifically, we implement a stepwise maximum-likelihood number density in each luminosity bin assuming a Poisson likelihood function. We estimate the uncertainty on the number density using a Markov Chain Monte Carlo (MCMC) approach, without imposing any prior on the number densities. At each iteration of the chain, we randomly draw $L_{\mathrm{bol}}$ values, within the uncertainty range, for each object, allowing individual sources to move between luminosity bins as the sampling proceeds. This procedure naturally accounts for both Poisson fluctuations and measurement uncertainties in $L_{\mathrm{bol}}$. The median of the posterior distribution is adopted as the best estimate of the luminosity function, while the 16th and 84th percentiles define the corresponding uncertainty range. The resulting number densities, per dex in bolometric luminosity, are reported in Table~\ref{tab:LF}.

We note that our luminosity function is robust to the main choices made in this study, namely: i) the photometric redshift choice, \texttt{CIGALE} vs \texttt{eazy} (\S~\ref{sec:z-det}), and ii) type-1+type-2 vs type-2 alone \texttt{CIGALE} (\S~\ref{sec:CIGALE-setup}). We illustrate these findings in Appendix~\ref{sec:app_LF}. \tb{We additionally tested the impact of photometric-redshift uncertainties on the LF measurements. For sources lacking spectroscopic redshifts, we perturbed the redshift at each step of the MCMC chain using the corresponding \texttt{CIGALE} uncertainty, and rescaled the bolometric luminosity according to the change in luminosity distance squared. We find that the resulting luminosity functions are robust to these perturbations, with negligible differences and no measurable impact on the inferred obscured fractions.}

\subsection{Parametric approach}\label{sec:LF_par}
To ensure continuous bolometric luminosity coverage in our luminosity function, necessary for meaningful comparison with existing literature (\S~\ref{Sec:results}), we also fit our results with a double power law, following standard practice:

\begin{equation}
    \phi(L)=\dfrac{{\rm d}n}{{\rm d}\log{L}}=\dfrac{\phi_{\ast}}{(L/L_{\ast})^{\gamma_1}+(L/L_{\ast})^{\gamma_2}},
    \label{eq:bPL}
\end{equation}
where $\phi_{\ast}$ is the comoving number density normalization, $L_{\ast}$ is the break luminosity, and $\gamma_1$ and $\gamma_2$ are the faint-end and bright-end slopes respectively.

Therefore, there are four free parameters in Eq.~\ref{eq:bPL}. However, given the limited data at the bright end ($L_\mathrm{bol} \gtrsim 10^{47}\,\mathrm{erg\,s^{-1}}$, Fig.~\ref{fig:Lbol_z}), we fix the bright-end slope, $\gamma_2$, to the value reported in \cite{Lacy2015}, $\gamma_2 = 2.48$. This choice is well motivated, as \cite{Lacy2015} employed a similar AGN selection function but extended to higher luminosities owing to their larger survey area. A caveat of this approach, however, is the lack of high-redshift ($z \gtrsim 3$) constraints. Nevertheless, \cite{Lacy2015} found no significant redshift evolution in the bright-end slope, and their $\gamma_2$ values are consistent with those of \cite{Shen2020} at high redshift, who fit a more homogeneous AGN sample. Adopting the \cite{Lacy2015} value for $\gamma_2$ is thus a reasonable choice and we note that modest deviations from this value have negligible impact on our results, as the luminosity function is primarily constrained by the more numerous, lower-luminosity AGN ($L_\mathrm{bol} \lesssim 10^{47}\,\mathrm{erg\,s^{-1}}$) that dominate our sample.

For the fit, we follow a similar approach to \cite{Finkelstein2015}. Namely, we compute the likelihood that the number of observed AGN in a given luminosity bin is equal to the predicted number for a given set of parameters. Instead of performing a grid-based search, we use MCMC to span the parameter space, and compute the uncertainties in the three parameters we fit. We use the same luminosity bins as in \S~\ref{sec:LF_non-par} (see also Table~\ref{tab:LF}). We model the probability distribution as Poissonian, due to the small-number statistics regime of our sample. For each redshift bin, we perform 10 independent MCMC chains, with 10000 steps each.

Lastly, we report our results from both the parametric and the non-parametric approach in Table~\ref{tab:LF}.

\begin{table*}
\centering
\caption{AGN bolometric luminosity functions in redshift bins, along with the best fit parameters for the double power-law (Eq.~\ref{eq:bPL}), \tb{as well as the corresponding completeness values, $C(L_{\mathrm{bol}},z)$, for both AGN ($\fracA > 0.5$), and mixed sources ($ 0.1 < \fracA <0.5$). The LF values reported in this table already include the completeness corrections.}}
\label{tab:LF}
\hspace*{-1.5cm}
\begin{tabular}{cccccc}
\toprule
Redshift  & Luminosity & \tb{Completeness - mixed} & \tb{Completeness - AGN} & $\Phi$  & Best-fit params\\
bin & bin (log) & \tb{\%} & \tb{\%} & $(\mathrm{Mpc}^{-3}\,\mathrm{dex}^{-1})$ & $\gamma_1, \rm{log}\phi_\star, \rm{log} L_\star$\\
\midrule
\multirow{6}{*}{$0.5<z<1.5$}
& 44-44.4  & \tb{96.92} & \tb{97.50} & $1.87^{+0.51}_{-0.43}\times10^{-4}$  &  \\
& 44.4-44.8& \tb{97.97} & \tb{98.02} & $1.02^{+0.36}_{-0.29}\times10^{-4}$  &$\gamma_1 = 0.73^{+0.15}_{-0.20}$\\
& 44.8-45.2  & \tb{98.58} & \tb{98.58} & $5.81^{+2.91}_{-2.14}\times10^{-5}$ &$\rm{log}\phi_\star = -4.73^{+0.59}_{-1.31} $ \\
& 45.2-45.6  & \tb{98.58} & \tb{98.58} & $1.44^{+1.55}_{-0.96}\times10^{-5}$  & $\rm{log}L_\star = 45.50^{+1.40}_{-0.48} $ \\
& 45.6 - 46  & \tb{98.58} & \tb{98.58} & $<4.34\times10^{-6}$    &\\
\midrule
\multirow{7}{*}{$1.5<z<2.5$}
& 44-44.4  & \tb{93.42} & \tb{93.46} & $1.39^{+0.39}_{-0.32}\times10^{-4}$ &\\
& 44.4-44.8& \tb{94.28} & \tb{94.46} & $1.04^{+0.30}_{-0.27}\times10^{-4}$   &\\
& 44.8-45.2  & \tb{95.90} & \tb{96.17} & $5.23^{+2.22}_{-1.77}\times10^{-5}$ & $\gamma_1 = 0.44^{+0.08}_{-0.09}$\\
& 45.2-45.6  & \tb{97.68} & \tb{97.51} & $4.77^{+2.06}_{-1.53}\times10^{-5}$ & $\rm{log}\phi_\star = -5.00^{+0.34}_{-0.46}$\\
& 45.6-46  & \tb{97.31} & \tb{98.45} & $2.13^{+1.52}_{-1.04}\times10^{-5}$ & $\rm{log}L_\star = 46.64^{+0.74}_{-0.43} $\\
& 46-46.4 & \tb{97.46} & \tb{98.35} & $1.83^{+1.37}_{-0.92}\times10^{-5}$  &\\
& 46.4-46.8 & \tb{96.97} & \tb{98.44} & $1.01^{+0.97}_{-0.59}\times10^{-5}$ &\\
\midrule
\multirow{6}{*}{$2.5<z<3.5$}
& 44.4-44.8& \tb{92.52} & \tb{92.87} & $1.45^{+0.37}_{-0.34}\times10^{-4}$ &\\
& 44.8-45.2  & \tb{93.21} & \tb{93.29} & $1.25^{+0.32}_{-0.29}\times10^{-4}$ & $\gamma_1 = 0.39^{+0.17}_{-0.19}$\\
& 45.2-45.6  & \tb{93.40} & \tb{93.77} & $8.44^{+2.72}_{-2.42}\times10^{-5}$ & $\rm{log}\phi_\star = -4.39^{+0.29}_{-0.50}$\\
& 45.6-46  & \tb{95.32} & \tb{95.40} & $4.38^{+1.97}_{-1.52}\times10^{-5}$ & $\rm{log}L_\star = 46.11^{+0.52}_{-0.26}$\\
& 46-46.4 & \tb{97.14} & \tb{96.95} & $2.21^{+1.30}_{-1.00}\times10^{-5}$ &\\
& 46.4-46.8 & \tb{97.88} & \tb{98.22} & $<3.90\times10^{-6}$ &\\
\midrule
\multirow{5}{*}{$3.5<z<4.5$}
& 44.8-45.2  & \tb{92.26} & \tb{92.52} & $6.44^{+2.99}_{-2.21}\times10^{-5}$ & \\
& 45.2-45.6  & \tb{92.71} & \tb{92.94} & $6.78^{+2.78}_{-2.24}\times10^{-5}$ & $\gamma_1 = 0.30^{+0.25}_{-0.20}$\\
& 45.6-46  & \tb{93.20} & \tb{92.92} & $4.79^{+2.34}_{-1.79}\times10^{-5}$  & $\rm{log}\phi_\star =-4.50^{+0.23}_{-0.40} $\\
& 46-46.4 & \tb{93.91} & \tb{93.59} & $1.93^{+1.48}_{-1.02}\times10^{-5}$  &$\rm{log}L_\star =46.20^{+0.39}_{-0.25} $\\
& 46.4 - 46.8 & \tb{96.25} & \tb{94.56} & $<4.84\times10^{-6}$  &\\
\midrule
\multirow{4}{*}{$4.5<z<6$}
& 44.2-44.9  & \tb{21.05} & \tb{11.94} & $2.50^{+2.69}_{-1.65}\times10^{-5}$ & \\
& 44.9-45.6  & \tb{78.45} & \tb{73.51} & $2.05^{+1.10}_{-0.87}\times10^{-5}$  & $\gamma_1 = 0.22^{+0.23}_{-0.15} $\\
& 45.6-46.3  & \tb{92.55} & \tb{92.67} & $1.77^{+0.90}_{-0.68}\times10^{-5}$  & $\rm{log}\phi_\star = -4.99^{+0.37}_{-0.56}$\\
& 46.3-47.0 & \tb{93.76} & \tb{93.15} & $<3.15\times10^{-6}$ & $\rm{log}L_\star =46.40^{+0.62}_{-0.44}  $\\

\bottomrule
\end{tabular}
\end{table*}

\section{Results}\label{Sec:results}
In this section, we use the luminosity functions described in \S~\ref{Sec:LF} to quantify the obscured AGN population missed by optical/UV studies. We first qualitatively show how our luminosity function, including both obscured and unobscured sources, compares with the type-1 UV-bright quasar luminosity function (QLF) reported in \cite{Kulkarni2019}. To estimate the obscured AGN fraction in each luminosity and redshift bin (\S~\ref{sect:f_obsc}), we integrate our MEOW/MIR luminosity function to determine the total AGN abundance (obscured + unobscured), and independently integrate the QLF for the unobscured population (see also Fig.~\ref{fig:roadmap}).

\subsection{Luminosity functions: MEOW AGN vs type-1 quasars}\label{sect:LF_MEOW_type1}

\begin{figure*}
    \includegraphics[width=\textwidth]{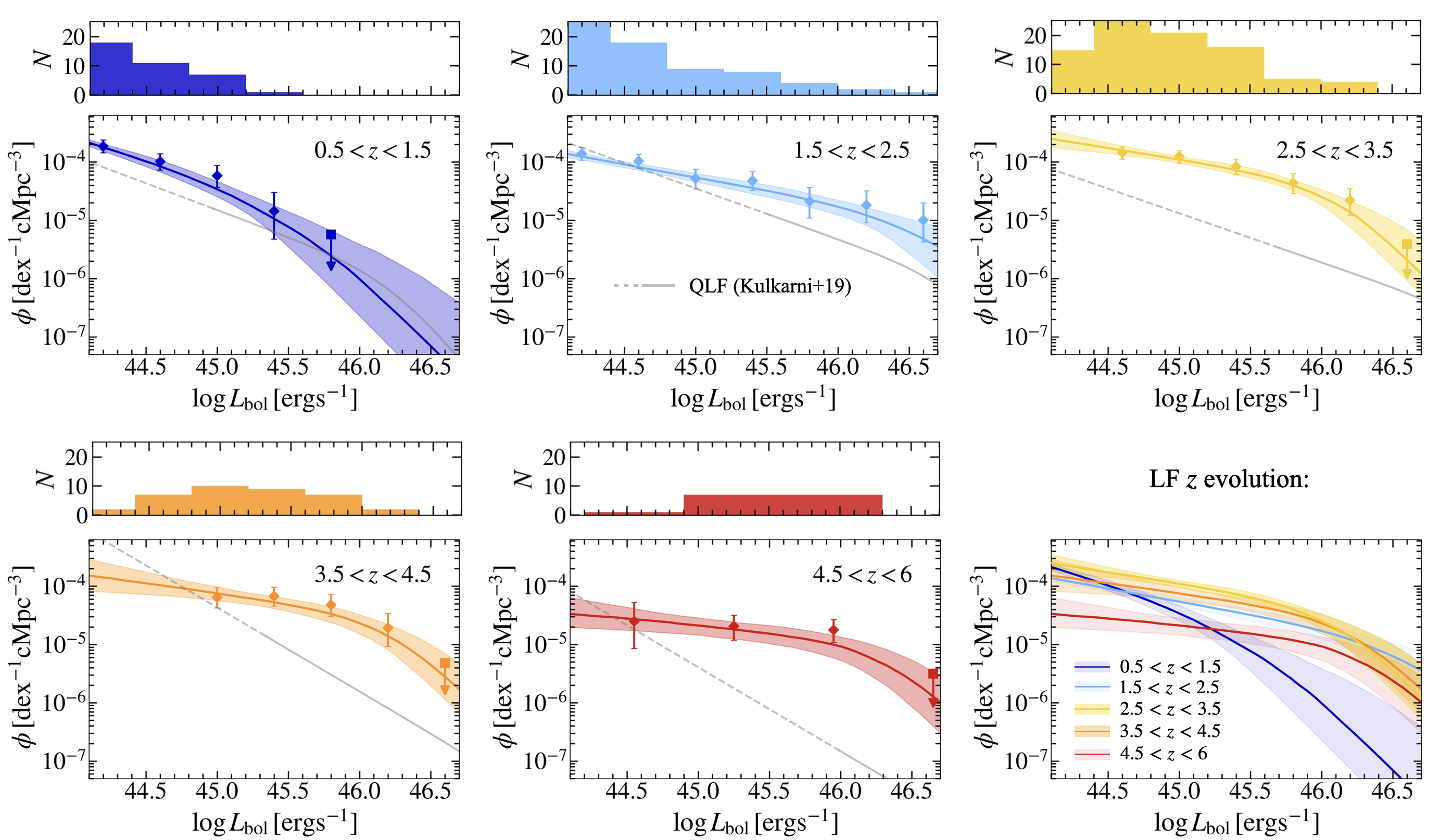}
    \caption{\textbf{AGN luminosity functions, along with the double power-law fits, color-coded by redshift.} The shaded regions represent the 16th-84th percentile of the MCMC chains (see \S~\ref{sec:LF_par}). For comparison, we show the luminous type-1 QLF from \cite{Kulkarni2019}, in grey, with the solid line representing the fits to the data, and the dashed line the extrapolations. We also show the AGN luminosity distributions in the histograms. We further quantify the excess of obscured AGN in our sample in Fig.~\ref{fig:fobsc}.}
    \label{fig:LF_all}

\end{figure*} 

We show both the non-parametric luminosity function (\S~\ref{sec:LF_non-par}), and the double power law fits (\S~\ref{sec:LF_par}), along with the corresponding uncertainties, in Figure~\ref{fig:LF_all}. \tb{To illustrate the redshift evolution more clearly, we combine all luminosity functions in the bottom-right panel of Fig.~\ref{fig:LF_all}.} We notice a smooth LF evolution between our redshift bins, with overall small uncertainties, apart from the high-luminosity end, due to small-number statistics. \tb{In the lowest-redshift bin, $0.5 < z < 1.5$, the luminosity function declines steeply at the bright end. By contrast, the higher-redshift luminosity functions turn over at higher luminosities, indicating a relatively larger abundance of luminous AGN. At lower luminosities, the luminosity functions become progressively flatter with increasing redshift, while the $4.5 < z < 6$ bin shows a substantially lower normalization. However, this low-luminosity regime is strongly affected by incompleteness (\S~\ref{sec:completeness}).} 

We compare our results to the ``Model 1" QLF reported in \cite{Kulkarni2019}, computed at the median redshift in each of our five redshift bins. \tb{The \cite{Kulkarni2019} sample consists exclusively of spectroscopically confirmed, UV-bright type-1 quasars extending to $z \lesssim 6$, making it well suited for estimating the obscured fractions of our mid-IR-selected AGN (see also \S~\ref{sec:LF_mess}).} We note that the QLF in \cite{Kulkarni2019} is originally reported in UV-magnitudes, $M_\mathrm{1450}$, which we convert to bolometric luminosity using the best fit of the \cite{Runnoe2012} relation. The uncertainties in the conversion introduce an additional source of error into our results that cannot be explicitly accounted for, but are expected to be subdominant at the population level. 

Overall, our results lie above the type-1 UV-bright QLF reported in \cite{Kulkarni2019}, in the regions with data coverage in \cite{Kulkarni2019} (i.e., not extrapolated). The excess of moderately luminous AGN in our sample suggests a substantial contribution from obscured growth that is not captured in UV-bright quasar surveys, which we will quantify in the next subsection. The accompanying number-count histograms reinforce this picture: while the bright end remains sparsely populated, a significant population of lower-luminosity AGN is detected out to $z \approx 6$, with a much higher abundance than the type-1 UV-bright quasars. Taken together, these results highlight the importance of mid-IR selection for tracing the full demographic of black hole growth across cosmic time, especially in regimes where dust obscuration is expected to be common (see also \S~\ref{sec:comparison})

\subsection{Obscured fraction}\label{sect:f_obsc}
\begin{figure*}
    \includegraphics[width=\textwidth]{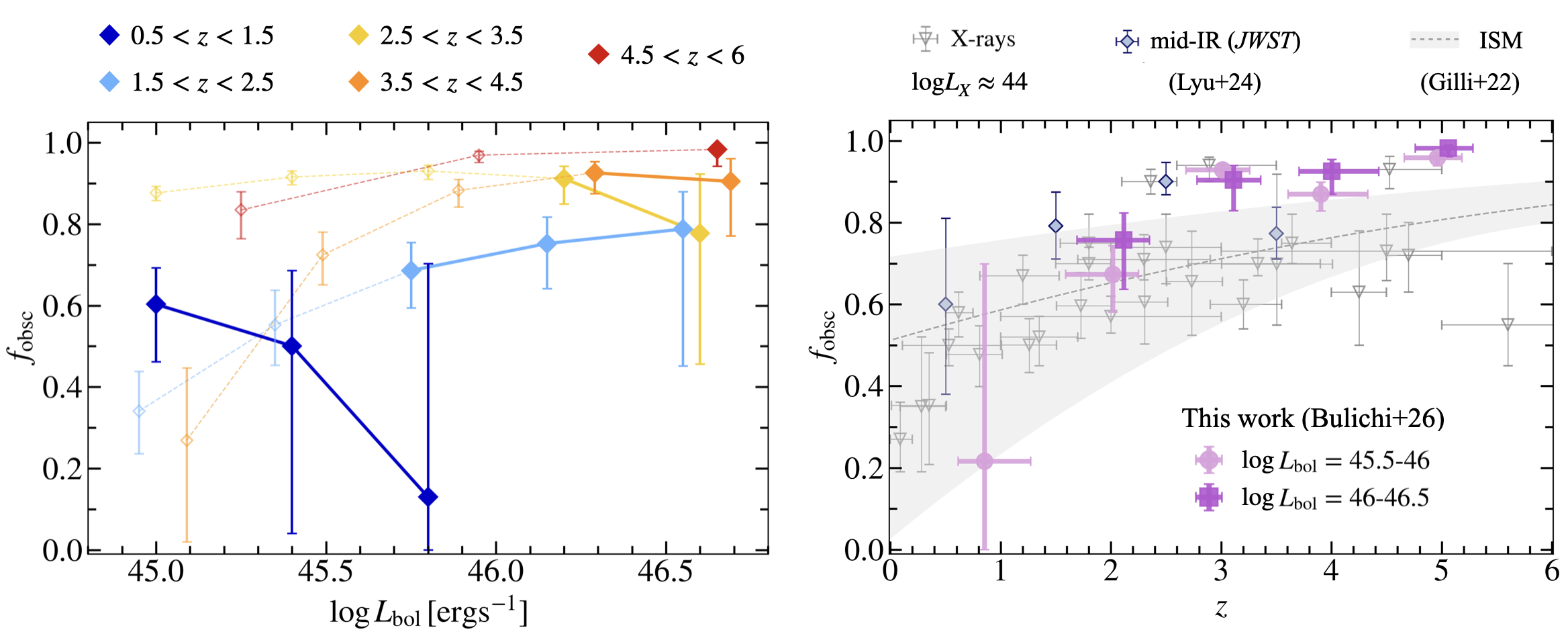}
    \caption{\textbf{Obscured fraction as function of luminosity, in five redshift bins (left), and as a function of redshift (right).} On the left, we show the results that come from QLF extrapolations (see Fig.~\ref{fig:LF_all}) as empty points, connected by dashed lines, and the rest of the data as solid points connected by solid lines. We color code the data by redshift, and use the same luminosity bins as in Fig.~\ref{fig:LF_all}. On the right, we show the redshift evolution across our faint sample (log $L_\mathrm{bol} = 45.5-46 \, \mathrm{erg s^{-1}}$; light purple), and bright sample (log $L_\mathrm{bol} = 46-46.5 \, \mathrm{erg s^{-1}}$; dark purple). The horizontal error bars indicate the 16th-84th percentile of the redshift distributions for the sources in each redshift bin. We overplot the X-ray constraints from \cite{Burlon2011,Iwasawa2012,Ueda2014,Buchner2015,Liu2017,Zappacosta2018,Vito2018,Iwasawa2020,Peca2023,Pouliasis2024,Barlow-Hall2025}, with 2-10 keV luminosities $\mathrm{log}\,L_X \approx 44 \,\mathrm{erg s^{-1}}$ as grey points; the ALMA constraints on $f_\mathrm{obsc}$ from the ISM column density \citep{Gilli2022}, as the grey band, with the dashed line showing the median predictions, as well as early \textit{JWST} results \citep{Lyu2024} as blue diamonds, indicating the optically obscured AGN in their sample.}
    \label{fig:fobsc}

\end{figure*}

\begin{table}
\centering
\caption{Obscured fractions in redshift bins, split in two luminosity ranges: $\mathrm{log}L_\mathrm{bol} = 45.5-46\,[\mathrm{erg\,s^{-1}}]$ (faint) and $\mathrm{log}L_\mathrm{bol} = 46-46.5\,[\mathrm{erg\,s^{-1}}]$ (bright), \tb{for all three models of \cite{Kulkarni2019}: Model 1 (fiducial), shown in the right panel of Fig.~\ref{fig:fobsc}, as well as Models 2 and 3.}}
\label{tab:fobsc}
\makebox[\textwidth][c]{\hspace*{-10.5cm}
\begin{tabular}{ccc}
\toprule
Redshift  & $f_\mathrm{obsc}$ (faint) & $f_\mathrm{obsc}$ (bright)   \\
bin & $\mathrm{log}L_\mathrm{bol} = 45.5-46$ & $\mathrm{log}L_\mathrm{bol} = 46-46.5$  \\
\midrule
$0.5<z<1.5$ & \textbf{$0.216^{+0.482}_{-0.216}$ (fiducial)} &  -- \\
& \tb{\hspace*{-1cm} K19 Model 2: $0.213^{+0.484}_{-0.213}$ } & -- \\ & \tb{\hspace*{-1cm}K19 Model 3: $0.216^{+0.483}_{-0.216}$} & -- \\
\midrule
$1.5<z<2.5$ & $0.673^{+0.070}_{-0.092}$ \textbf{(fiducial)} & $0.757^{+0.066}_{-0.120}$\textbf{(fiducial)} \\
& \tb{\hspace*{-1cm} K19 Model 2: $0.640^{+0.077}_{-0.102}$ } & \tb{$0.755^{+0.067}_{-0.121}$}  \\ 
& \tb{\hspace*{-1cm}K19 Model 3: $0.663^{+0.072}_{-0.096}$} & \tb{$0.767^{+0.064}_{-0.115}$} \\
\midrule
$2.5<z<3.5$ & $0.929^{+0.013}_{-0.019}$\textbf{(fiducial)} & $0.904^{+0.034}_{-0.076}$\textbf{(fiducial)}  \\
& \tb{\hspace*{-1cm} K19 Model 2: $0.764^{+0.046}_{-0.066}$ } & \tb{$0.823^{+0.064}_{-0.141}$}  \\ 
& \tb{\hspace*{-1cm}K19 Model 3: $0.904^{+0.019}_{-0.027}$} & \tb{$0.896^{+0.038}_{-0.083}$} \\
\midrule
$3.5<z<4.5$ & $0.870^{+0.028}_{-0.043}$\textbf{(fiducial)}  & $0.926^{+0.028}_{-0.057}$ \textbf{(fiducial)} \\
& \tb{\hspace*{-1cm} K19 Model 2: $0.931^{+0.015}_{-0.022}$ } & \tb{$0.958^{+0.016}_{-0.032}$}  \\ 
& \tb{\hspace*{-1cm}K19 Model 3: $0.969^{+0.007}_{-0.010}$} & \tb{$0.974^{+0.010}_{-0.020}$} \\
\midrule
$4.5<z<6$ & $0.958^{+0.012}_{-0.020}$\textbf{(fiducial)}   & $0.983^{+0.007}_{-0.022}$\textbf{(fiducial)}   \\
& \tb{\hspace*{-1cm} K19 Model 2: $0.974^{+0.007}_{-0.012}$ } & \tb{$0.989^{+0.005}_{-0.014}$}  \\ 
& \tb{\hspace*{-1cm}K19 Model 3: $0.985^{+0.004}_{-0.007}$} & \tb{$0.992^{+0.003}_{-0.010}$} \\
\bottomrule
\end{tabular}
}
\end{table}

We quantify the obscured fraction using the luminosity functions of our MEOW sample, compared to the type-1 UV-bright QLF reported in \cite{Kulkarni2019} (see \S~\ref{sect:LF_MEOW_type1}). Specifically, we define the obscured fraction as the abundance of obscured sources over the total AGN number density:

\begin{equation*}
    f_\mathrm{obsc} = \frac{n_\mathrm{obsc}}{n_\mathrm{tot}} = 1- \frac{n_\mathrm{unobsc}}{n_\mathrm{tot}}
\end{equation*}

\setcounter{footnote}{0}

\noindent where $n_\mathrm{unobsc} = \int \phi_\mathrm{QLF}(L) , \mathrm{d}L$ is the number density of unobscured, UV-bright type-1 AGN, obtained by integrating the luminosity function from \cite{Kulkarni2019}. In contrast, $n_\mathrm{tot} = \int \phi_\mathrm{MIR}(L) , \mathrm{d}L$ is the total AGN number density, obtained by integrating our MIR parametric LF (\S~\ref{sec:LF_par}; see also Fig.~\ref{fig:roadmap})\footnote{\tb{Throughout this study, our inferred obscured fractions refer strictly to UV obscuration, i.e., the fraction of AGN missed by UV selection.}}. \tb{This total includes both obscured and unobscured AGN, since the MEOW selection is sensitive to hot-dust emission irrespective of orientation, as demonstrated in \S~\ref{sect:AGN_identification}.}

\tb{As discussed further in \S~\ref{sec:LF_mess}, we adopt the \cite{Kulkarni2019} QLF as the most appropriate comparison for this study, as it specifically represents UV-bright, type-1 AGN/quasars and provides reasonable coverage across the redshift and luminosity ranges probed here. Generally, the low luminosity regimes (i.e., $L_\mathrm{bol} < 10^{45}\,\mathrm{erg \, s^{-1}}$) do not overlap with our data, as indicated by the dashed lines in Figure~\ref{fig:LF_all}; however, we do not draw conclusions from the obscured fractions inferred in these regimes. Likewise, as discussed in \S~\ref{sec:completeness}, our low-luminosity measurements in the highest-redshift bins are strongly affected by completeness corrections and fall in a luminosity range not covered by \cite{Kulkarni2019}. We therefore exclude these measurements from the final obscured fraction calculations.}

We show the obscured fraction evolution as a function of luminosity, for each redshift bin, as well as a function of redshift in Figure~\ref{fig:fobsc}. We denote the results coming from the extrapolation in the \cite{Kulkarni2019} QLF with a dashed line, and empty points, and the rest with solid points connected by solid lines. We estimate the uncertainties by sampling the luminosity-function MCMC chains (\S~\ref{sec:LF_par}) and computing the 16th-84th percentile range of the resulting obscured fractions.

The obscured AGN fraction ($f_\mathrm{obsc}$) in the MEOW sample displays a \tb{tentative} luminosity dependence at low redshift, with the lowest redshift bin hinting at decreasing obscuration toward higher luminosities, albeit \tb{not statistically significant, due to the large uncertainties}. This \tb{tentative} behavior is broadly consistent with a ``receding torus'' model (e.g., \citealt{Lawrence1991,Simpson2005,Gilli2007,Lusso2013}), which predicts that the high luminosities reduce the dust covering factor. However, the trend reverses in the higher-redshift bins, hinting at an evolving obscuration scenario (e.g., \citealt{Sanders1988,Hopkins2008}, see \S~\ref{sec:comparison}).


We additionally notice an increase in obscuration with redshift for the luminosity ranges covered by the \cite{Kulkarni2019} QLF, explored further in the right panel of Fig.~\ref{fig:fobsc}, \tb{and also included in Table~\ref{tab:fobsc}}. We split our sources in two luminosity bins: ${\rm{log}} L_{\rm{bol}} = 45.5-46\,\rm{erg\,s^{-1}}$, and ${\rm{log}} L_{\rm{bol}} = 46-46.5\,\rm{erg\,s^{-1}}$, to differentiate between fainter and brighter sources while maintaining comparable luminosity coverage with the UV-bright quasars in \cite{Kulkarni2019}, as well as meaningful comparisons with X-ray studies \citep{Burlon2011,Iwasawa2012,Ueda2014,Buchner2015,Liu2017,Zappacosta2018,Vito2018,Iwasawa2020,Pouliasis2024,Barlow-Hall2025}. 

The redshift evolution we find aligns qualitatively with X-ray studies, but our mid-IR selection generally identifies an even larger proportion of highly obscured systems, and is above the indirect ISM constraints from \cite{Gilli2022} at high redshift. When compared to the early \textit{JWST} results of \cite{Lyu2024}, we find similarly high levels of optical obscuration over the overlapping redshift range. These results suggest that much of the black hole growth \tb{(up to 98-99\%)} is deeply embedded, with heavy dust columns capable of concealing luminous accretion from optical and even hard X-ray surveys (see also \S~\ref{sec:comparison}).


\section{Discussion}\label{Sec:discussion}
In this section, we provide a detailed interpretation of the results presented in \S~\ref{Sec:results}. We first compare our findings with previously reported luminosity functions (\S~\ref{sec:LF_mess}) and motivate our choice of the QLF used in the obscured-fraction calculations (\S~\ref{sect:f_obsc}). We then place our sample in the context of prior multi-wavelength studies, as well as existing literature models, highlighting the newly identified obscured AGN uncovered in this work (\S~\ref{sec:comparison}). Finally, we interpret the inferred obscured fractions in the broader context of supermassive black hole growth (\S~\ref{sec:SMBH_growth}).

\subsection{Choice of quasar luminosity function}
\label{sec:LF_mess}

\begin{figure}
    \includegraphics[width=\columnwidth]{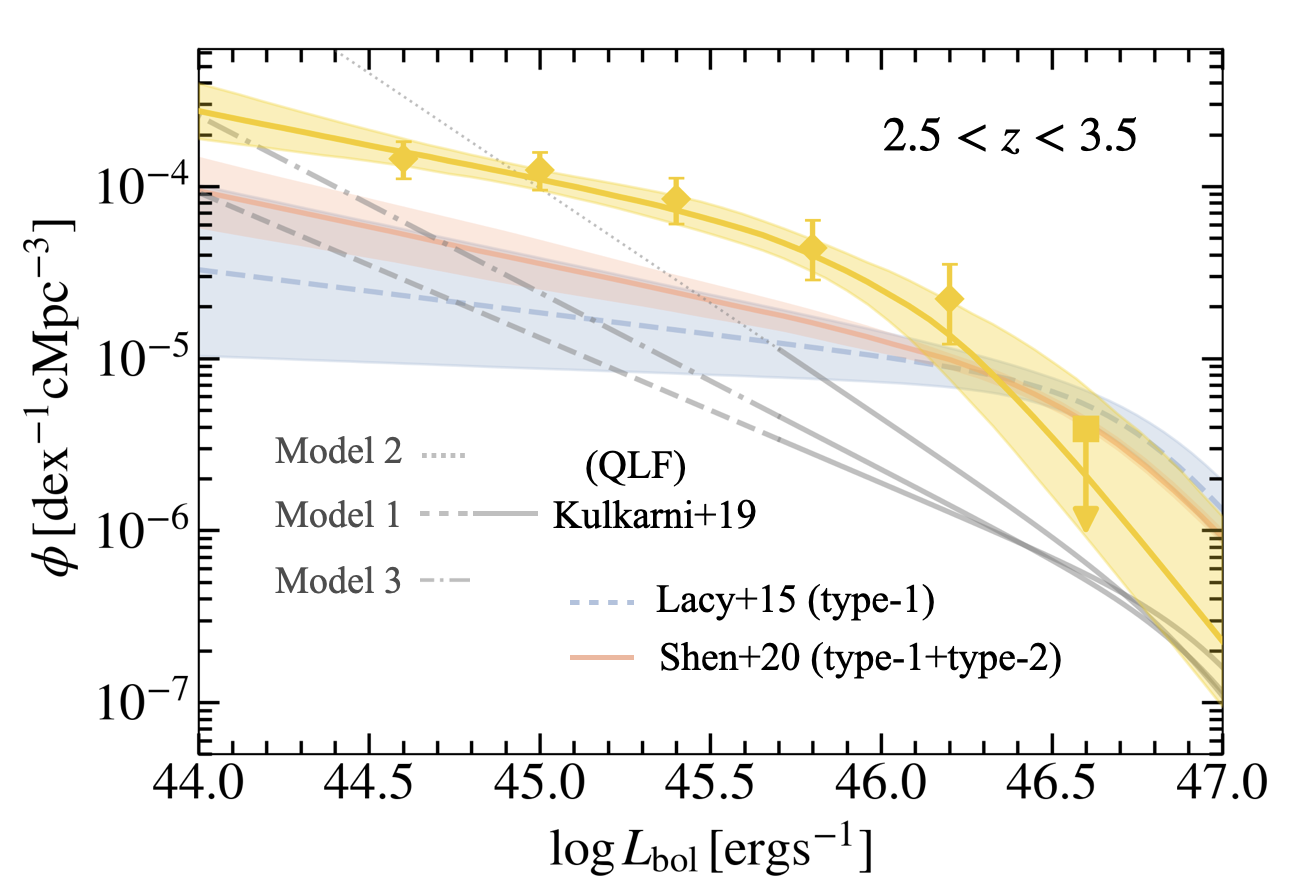}
    \caption{\textbf{AGN luminosity function, for our third redshift bin: $2.5 < z < 3.5$, in comparison with previous literature results.} For all the QLFs shown, we use the value at the median redshift in our distribution, and the shaded regions represent the errors in the fitting. The solid-dashed line is the \cite{Kulkarni2019} model, same as in Fig.~\ref{fig:LF_all}, and we additionally plot the other two models reported in \cite{Kulkarni2019}, which account for different selection effects (see text), as well as the QLFs reported in \cite{Lacy2015} (type-1) and \cite{Shen2020} (type-1+type-2). Using the same convention as in Fig.~\ref{fig:LF_all}, the dashed lines show the regions where the LFs are extrapolated.}
    \label{fig:LF_mess}

\end{figure} 

While the luminosity function is a powerful probe of structure formation and evolution \citep{Roberts2026}, its accuracy is still strongly limited by selection effects. As such, the findings of this study are highly sensitive to the comparison with the QLF used for the $f_\mathrm{obsc}$ calculations (\S~\ref{sect:f_obsc}).  Here, we investigate all models for the UV-selected QLFs reported in \cite{Kulkarni2019}, along with the results from the mid-IR selected AGN in \cite{Lacy2015} and the multi-wavelength (IR-to-X-ray) sample of \cite{Shen2020}, and show all of these results in Fig.~\ref{fig:LF_mess}, at $2.5 < z < 3.5$, $z_\mathrm{med} = 3.06$.

\cite{Kulkarni2019} provide robust redshift and completeness estimates, applying homogeneous assumptions for the cosmology and the intrinsic AGN SED, addressing the large methodological scatter present in preceding work.
We already showed Model 1 in \S~\ref{Sec:LF}, which contains all the datasets compiled in the paper, and models the faint end slope as a double power law. Their Model 2 excludes the results with only approximate selection functions, while Model 3 further modifies the faint-end slope as a linear fit in $(1+z)$. Although these models show reasonable agreement where data exist, they diverge significantly toward the low-luminosity regime at high redshift, where there are no observational data points. We decided to use Model 1 as the baseline of this work, as it is the most complete, but note that the results are sensitive to this choice, especially in the extrapolated regions, and thus the systematic uncertainties in the low luminosity regime are much higher than the LF fitting uncertainties quoted in Table~\ref{tab:fobsc}. \tb{We therefore do not draw conclusions in luminosity ranges where the fits from \cite{Kulkarni2019} are extrapolated. In the luminosity ranges directly covered by the \cite{Kulkarni2019} data, we report the corresponding obscured fractions for Models 2 and 3 in Table~\ref{tab:fobsc}. Overall, the results from the three models agree within $1\sigma$, with the exception of the $2.5 < z < 3.5$ bin, where the \cite{Kulkarni2019} models differ most substantially, as also shown in Figure~\ref{fig:LF_mess}. While we consider Model 1 to be the most suitable choice for this work, we caution that the obscured fraction at $z \approx 3$ is sensitive to this model choice, as reflected in Table~\ref{tab:fobsc}. The results in the other redshift bins remain robust to this assumption.}

\citet{Lacy2015} selects the AGN sample in mid-IR with \textit{Spitzer}, similar to this work. Using spectroscopic follow-up, they differentiate between type-1 and type-2 based on the broad emission lines: objects with a broad component are classified as type-1. While the initial selection effects are similar to those used in this paper, the \cite{Lacy2015} fits are limited in redshift $z \lesssim 3$, and luminosity $L_\mathrm{bol} \gtrsim 10^{47}\,\mathrm{erg s^{-1}}$, thus not overlapping with the data in our sample. Thus, we did not use the \cite{Lacy2015} type-1 luminosity function to quantify obscuration, but note that their results predict larger $\phi$ values than \cite{Kulkarni2019}. We speculate that this might result from the small-number statistics in \cite{Lacy2015}, as the number of sources is significantly lower than in \cite{Kulkarni2019}, as well as the different classifications as type-1. We however use the \cite{Lacy2015} results for the bright end LF constraints (\S~\ref{sec:LF_par}), as their methodology for the whole AGN population (type-1+type-2) aligns closely with ours.
 
Lastly, while \citet{Shen2020} include both obscured and unobscured AGN, their high-redshift obscured population is dominated by X-ray-selected sources. As a result, this sample may be affected by survey incompleteness, uncertainties in X-ray-to-UV bolometric corrections, and biases against Compton-thick AGN. Consequently, \citet{Shen2020} exhibit a behavior similar to \citet{Lacy2015} at high redshift, where only a small fraction of obscured AGN are represented. The fact that our luminosity function lies above that of \citet{Shen2020} underscores the effectiveness of mid-IR selection in identifying obscured AGN that are otherwise missed in UV and X-ray surveys, an effect that is particularly pronounced at lower luminosities ($L_\mathrm{bol} \lesssim 10^{46}\,\mathrm{erg\,s^{-1}}$). We note, however, that at lower redshift, our luminosity function shows good agreement with \citet{Shen2020}, where the inclusion of mid-IR-selected AGN in \cite{Shen2020} leads to a more complete census of both obscured and unobscured populations.

We further quantify the missing obscured AGN compared to previous results and models in the next subsection, and conclude by discussing the implications of our inferred obscured fraction to early SMBH growth (\S~\ref{sec:SMBH_growth}).

\subsection{Comparison with literature AGN samples}
\label{sec:comparison}
\begin{figure*}
    \includegraphics[width=\textwidth]{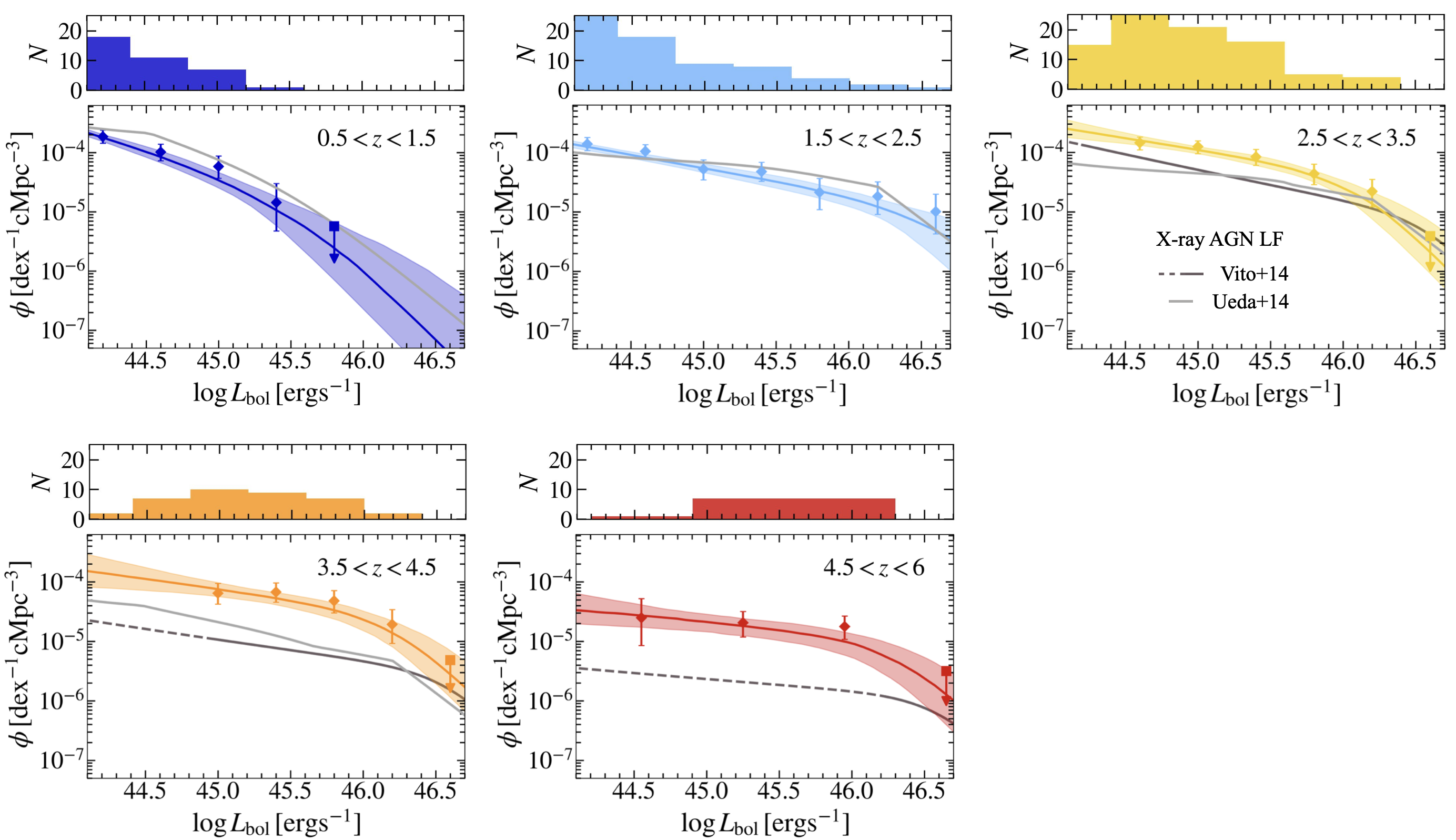}
    \caption{\textbf{\tb{Comparison of the bolometric AGN luminosity function for MEOW-selected AGN (colored lines with shaded regions), and X-ray selected AGN (grey lines, from \citealt{Ueda2014} and \citealt{Vito2014}).}} \tb{Similar to Fig.~\ref{fig:LF_all}, the dashed lines mark the luminosity range where the LFs are extrapolated. At low redshift ($z < 3$), our LFs agree well with those inferred from X-ray selected AGN. At $z\gtrsim 3$, the higher AGN abundance compared to the X-ray selected AGN luminosity functions suggests the existence of an additional, heavily obscured AGN population not captured by X-ray surveys, with the excess increasing toward higher redshift.}}
    \label{fig:XLF}

\end{figure*} 

As shown in Fig.~\ref{fig:fobsc}, our inferred obscured fractions are generally higher than those derived from X-ray studies, at high redshift. This discrepancy can be partly attributed to the different methodology used to infer $f_\mathrm{obsc}$ in this work compared to X-ray-based analyses. However, an additional contributor is the presence of X-ray weak AGN in our sample, potentially arising from high dust covering fractions and/or large column densities (i.e., Compton-thick AGN). To investigate this scenario, we compare our luminosity functions with the luminosity functions of X-ray selected AGN, reported in \cite{Vito2014} \tb{and \cite{Ueda2014}, in Figure~\ref{fig:XLF}}, using their ``luminosity-dependent density evolution'' fit at the median redshift of our sample in each redshift bin. \tb{We include both X-ray-selected AGN luminosity functions because they probe complementary redshift and luminosity regimes, together providing the closest comparison to the parameter space covered by our sample. In both cases, the conversion between the X-ray luminosity $L_{2-10\,\rm{keV}}$ and bolometric luminosity was done using the relation reported in \cite{Hopkins2007}, and applying the corresponding transformation to the luminosity function.}

\tb{In the first two redshift bins, our luminosity functions agree, within the uncertainties, with the corresponding X-ray-selected AGN luminosity functions. This is consistent with the agreement between our inferred obscured fractions and those derived from X-ray studies over the same redshift range (Fig.~\ref{fig:fobsc}), suggesting that at $z \lesssim 3$, X-ray luminosity functions are less strongly affected by incompleteness and recover a substantial fraction of the underlying AGN population (e.g., \citealt{Guo2026,Mazzolari2026}). At higher redshifts, however, our mid-infrared selection identifies additional AGN that are not detected in X-rays, pointing to a population of highly obscured systems and/or sources with large gas columns that may not be fully accounted for in current X-ray selections. This is also reflected in our systematically higher obscured fractions relative to X-ray-based estimates at $z \gtrsim 3$ (Fig.~\ref{fig:fobsc}).}



Compared to early \textit{JWST} results \citep{Lyu2024}, we adopt a different definition to quantify obscuration. In \cite{Lyu2024}, obscuration is characterized through the AGN contribution to the total emission in the rest-frame 1-10~$\mu\mathrm{m}$ range, which, within the \texttt{CIGALE} SED-fitting framework implemented in this study, is strongly degenerate with the emission from the stellar component. As a result, a strictly direct comparison between the two approaches is not possible. Nevertheless, we find overall consistency with the obscuration levels reported in \cite{Lyu2024}. We additionally cross-match our sample with the AGN identified by \cite{Lyu2024}, finding good overlap in the luminosity regimes probed by both studies: \tb{Of the 90 SMILES sources above the MEOW detection limits, we recover 84. The six unrecovered sources are missed due either to the lack of F444W coverage, in two cases, or to differences in the deblending, in the remaining four. Since the latter effect is included in our completeness simulations, it is not expected to introduce systematic uncertainties in the luminosity function or obscured-fraction measurements. We} discuss further the implications of this comparison in Appendix~\ref{sec:app_Lyu}.




We also find that our inferred obscured fractions show a mild dependence on luminosity and a strong redshift dependence. In the lowest redshift bin, $0.5 < z < 1.5$ the obscured fraction goes down with luminosity, in agreement with ``receding torus'' models \citep{Lawrence1991,Simpson2005,Gilli2007,Lusso2013}, which predict that higher luminosities drive the inner radius of the torus outward, thereby reducing the dust covering fraction. At higher redshifts, however, we notice the opposite trend, with the caveat that this is mainly driven by the regions where the \cite{Kulkarni2019} QLF is extrapolated. This is in agreement with previous \textit{Spitzer} mid-IR studies (e.g., \citealt{Lacy2015}), which support evolutionary scenarios proposed by e.g., \cite{Sanders1988} and \cite{Hopkins2008}, in which quasars begin their activity in highly obscured environments. This is also supported by the increase in obscuration with redshift shown in Figure~\ref{fig:fobsc}, which agrees with earlier work in X-ray surveys (e.g., \citealt{Treister2006,Liu2017}). However, X-ray studies suggest that, at higher redshifts, obscuration is dominated by dust distributed on galactic scales rather than by the compact, hot torus traditionally invoked in AGN unification models (e.g., \citealt{Gilli2022,Trefoloni2025}). In contrast, we still notice a significant increase in obscuration with our MIRI-based analysis that primarily traces hot dust emission associated with AGN heating. This apparent contradiction may arise because many previous X-ray constraints were derived from lower-redshift samples ($z \lesssim 2$), where the physical conditions and dominant obscuration mechanisms could differ. Second, while the hot dust we detect is consistent with AGN heating, it does not necessarily require a classical torus geometry; it may instead originate from dust distributed more broadly, provided it is sufficiently heated by the central engine. Lastly, the rest-frame NIR emission from the hot torus will better penetrate the galaxy dust than the rest-UV and optical emission, making it a useful tracer in either case of obscuration (see e.g., \citealt{Silverman2023}).

Finally, we caution that the systematic uncertainties, especially due to the choice of the QLF, might be underestimated, thus affecting some obscured fraction estimates, predominantly at the low luminosity end (see \S~\ref{sect:LF_MEOW_type1}). However, the high luminosity end constraints from \cite{Kulkarni2019} converge to similar values, making the $f_\mathrm{obsc}$ estimations more robust in this regime. This strengthens our confidence in the obscured fraction at high redshift and high luminosity, discussed in more detail in the next subsection.

\subsection{Implications for SMBH growth}\label{sec:SMBH_growth}

As detailed in e.g., \cite{Eilers2018,Eilers2024}, the short UV-bright lifetimes inferred independently from ionizing signatures, i.e., proximity zones, Ly$\alpha$ nebulae and damping wings measurements \citep{Eilers2017,Eilers2018,Davies2019,Eilers2020,Andika2020,Morey2021,Satyavolu2023,Durovcikova2024,Durovcikova2025}, as well as population study duty cycles \citep{Arita2023,Eilers2024,Pizzati2024,Schindler2025,Huang2026}, can be alleviated if the early SMBH growth occurs in heavily dust-enshrouded environments that suppress the ionizing UV emission.

The results of our study point towards high obscured fractions, as high as $f_\mathrm{obsc} \approx 98-99\%$ in the highest redshift bin at $4.5 <z< 6$ \tb{(Table~\ref{tab:fobsc})}. This could alleviate some of the tension raised by the short UV-luminous duty cycles of order $\sim 0.1-1\%$ reported by recent studies at $z \gtrsim 6$ \citep{Eilers2024,Pizzati2024,Schindler2025,Huang2026}, as it implies that there is a large fraction of accreting obscured black holes. \tb{We caution, however, that duty-cycle studies generally probe higher luminosities ($L_\mathrm{bol} \sim 10^{47}\,\mathrm{erg\,s^{-1}}$) and higher redshifts ($z \gtrsim 6$) than those considered in this work. A direct comparison is therefore not currently possible. In the future however, as more photometric and spectroscopic \textit{JWST} datasets become available (e.g., \textit{JWST} \#6827), it will be possible to quantify the clustering and duty cycles of both obscured and unobscured AGN over a common $L_\mathrm{bol}$--$z$ parameter space, providing a more complete view of SMBH growth. At present, our results suggest that the high obscured fractions inferred at $z \approx 5$ and $L_\mathrm{bol} \sim 10^{45}-10^{46}\,\mathrm{erg\, s^{-1}}$ can help reconcile rapid SMBH growth, with up to $\approx 98-99\%$ of accretion occurring in dust-enshrouded phases missed by traditional optical/UV surveys. If similarly high obscured fractions persist at $z \gtrsim 6$ and at higher luminosities, they would provide an important pathway for explaining the rapid assembly of early SMBHs, despite the short luminous/quasar phases.}


\section{Conclusions}\label{Sec:conclusions}
We have presented a measurement of the obscured AGN fraction from $0.5 \lesssim z\lesssim 6$ based on the \textit{JWST}/MIRI MEOW sample in the two GOODS fields. Using full SED fitting to identify the AGN and to infer their bolometric luminosities, we construct luminosity functions of the MIR-selected AGN in five redshift bins and compare these to the type-1 UV-bright quasar luminosity function from \cite{Kulkarni2019}. 

Our results reveal a substantial population of dust-enshrouded AGN at early cosmic times. In particular, at high redshift ($z \gtrsim 4.5$) and high luminosity ($\log L_{\mathrm{bol}} \gtrsim 46\,\mathrm{erg\,s^{-1}}$), the estimated obscured fraction approaches $\approx 98-99\%$. 
While the detailed luminosity dependence of the obscured fraction remains sensitive to the choice of type-1 QLF at the faint end, the high-luminosity regime is comparatively well constrained, suggesting that the black hole accretion timescales could be up to two orders of magnitude larger than previous estimates based on proximity zones and clustering of UV-luminous quasars. Future measurements of quasar lifetimes from damping-wing signatures in faint quasars (JWST\#9180, PI: Hennawi) and from clustering analyses (JWST\#7519, PI: Arita) will be key to further constraining this interpretation.

Our estimates of the obscured AGN fraction exceed those derived in previous studies from X-ray-selected samples at $z\gtrsim 3$. This suggests that a significant fraction of SMBH growth at early times occurs in heavily dust-embedded environments, with a significant population of heavily obscured/Compton-thick AGN missed by X-ray selections towards higher redshifts.


Overall, these results highlight the indispensable role of mid-infrared selection in constructing a complete census of SMBH growth at high redshift. Future spectroscopic follow-up, alongside wider and deeper MIRI surveys, will be key to refining obscuration estimates and uncovering the physical conditions and environments that regulate early, dust-rich black hole growth. With forthcoming data from programs such as \textit{JWST}\#6827 (PI: Eilers), this work will serve as an important benchmark for placing tighter constraints on early SMBH assembly and for testing evolutionary scenarios in which black hole growth progresses from a heavily obscured phase to a later, more unobscured state.


\begin{acknowledgments}
The authors would like to thank \tb{the referee for a very constructive report that helped in improving the clarity of the paper. We also thank Allison Kirkpatrick,} Giovanni Mazzolari, Dominika \v{D}urov{\v{c}}{\'\i}kov{\'a}, Megan Masterson, John Silverman, Erini Lambrides and Johannes Buchner for helpful discussions. T-E.B. and G.C.K.L acknowledge support from NASA through STScI award JWST-GO-5407. Support for program \#5407 was provided by NASA through a grant from the Space Telescope Science Institute, which is operated by the Association of Universities for Research in Astronomy, Inc., under NASA contract NAS 5-03127. This work is based on observations made with the NASA/ESA/CSA James Webb Space Telescope. The data were obtained from the Mikulski Archive for Space Telescopes at the Space Telescope Science Institute, which is operated by the Association of Universities for Research in Astronomy, Inc., under NASA contract NAS 5-03127 for JWST. PGP-G acknowledges support from grant PID2022-139567NB-I00 funded by Spanish Ministerio de Ciencia, Innovaci\'on y Universidades MCIU/AEI/10.13039/501100011033,FEDER {\it Una manera de hacer Europa}.
\end{acknowledgments}

%

\software{astropy \citep{2013A&A...558A..33A,2018AJ....156..123A,2022ApJ...935..167A}, eazy \citep{eazy}, Source Extractor \citep{1996A&AS..117..393B}, Matplotlib\citep{hunter2007}, NumPy \citep{numpy}, SciPy \citep{jones2001}; JWST Calibration Pipeline \citep{bushouse_howard_2022_7429939}}.





\appendix

\section{Tabulated data}\label{sec:tab_data}
We provide the IDs, coordinates and physical properties of our AGN sample ($f_\mathrm{AGN}>0.1$) in Table~\ref{tab:data}, available in its entirety in machine-readable form in the online article.

\begin{table*}[h!]
    \centering
    \begin{tabular*}{\textwidth}{@{\extracolsep{\fill}} c c c c c c c @{}}
        \hline
        NIRCam ID & RA & DEC & $z$ & $z_\mathrm{type}$ & $f_\mathrm{AGN}$ & $\log L_\mathrm{bol}$ \\
        & [deg] & [deg] & & & & [$\mathrm{erg\,s^{-1}}$] \\
        \hline
        104301 & \tb{189.354997} & \tb{62.241363} & 0.1046 & phot & $0.90^{+0.02}_{-0.02}$ & $41.07^{+0.04}_{-0.05}$ \\
        103521 & \tb{189.354893} & \tb{62.250678} & 0.4577 & spec & $0.32^{+0.10}_{-0.10}$ & $42.35^{+0.13}_{-0.19}$ \\
        17959  & \tb{189.213667} & \tb{62.181003} & 1.2258 & spec & $0.83^{+0.11}_{-0.11}$ & $44.72^{+0.07}_{-0.08}$ \\
        6281   & \tb{189.184761} & \tb{62.248055} & 1.4870 & spec & $0.32^{+0.13}_{-0.13}$ & $44.84^{+0.16}_{-0.26}$ \\
        7843   & \tb{189.205762} & \tb{62.238238} & 3.1934 & phot & $0.52^{+0.16}_{-0.16}$ & $45.22^{+0.11}_{-0.15}$ \\
        \hline
    \end{tabular*}
    \caption{The table is available in a machine-readable format in the online article.}
    \label{tab:data}
\end{table*}

\section{Luminosity function dependence on modeling choices}\label{sec:app_LF}

\begin{figure*}[!h]
    \centering

    \begin{minipage}[t]{0.33\textwidth}
        \vspace{0pt}
        \includegraphics[width=\linewidth]{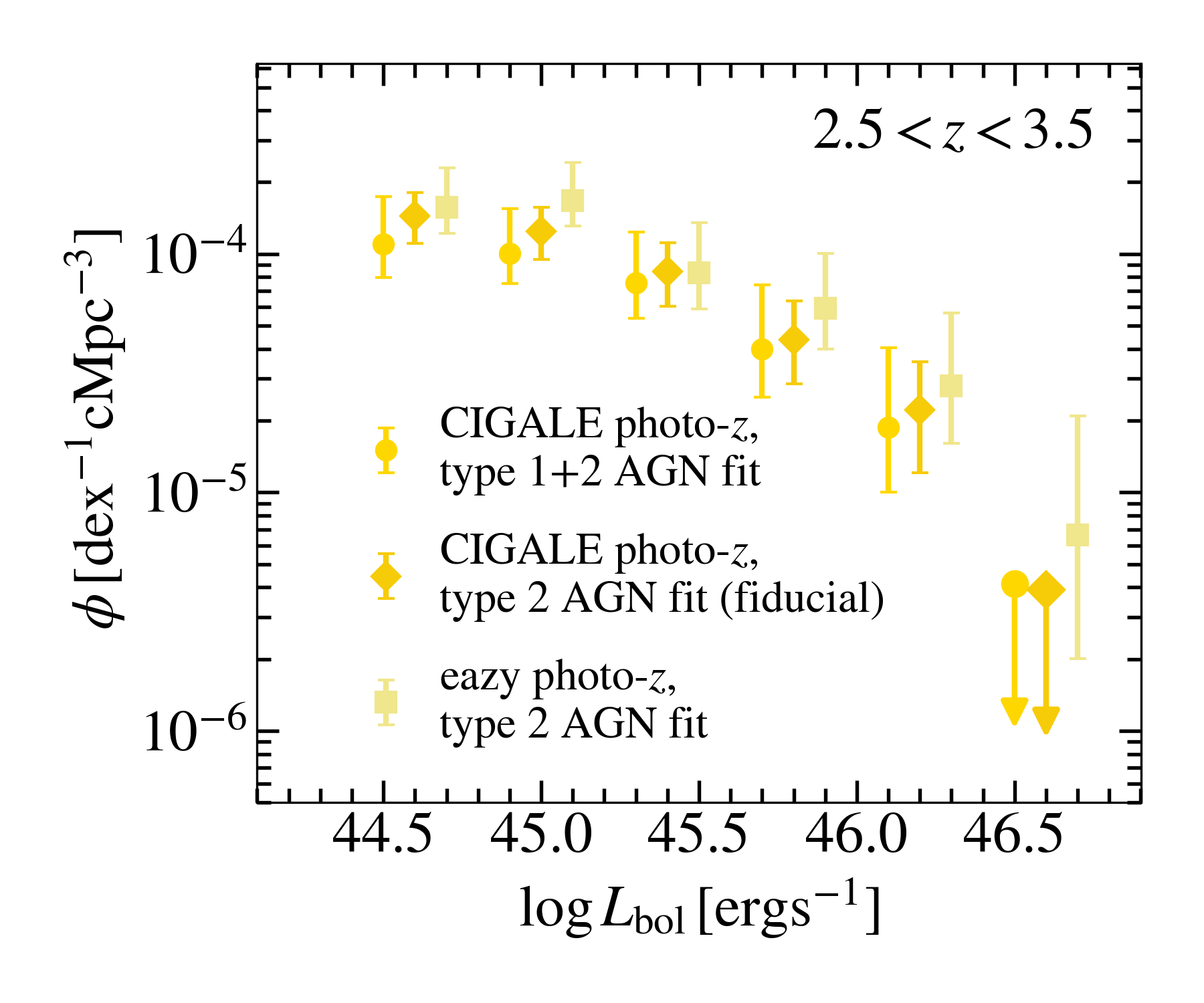}
    \end{minipage}
    \hfill
    \begin{minipage}[t]{0.65\textwidth}
        \vspace{0.2cm}
        \caption{\tb{\textbf{AGN luminosity function under different modeling assumptions.} The points are shifted horizontally for better visualization. The yellow diamonds show our fiducial model, which uses \texttt{CIGALE} photometric redshifts for sources without spectroscopic redshifts and models the AGN emission with a type-2 template only. The yellow circles also use \texttt{CIGALE} photometric redshifts, but additionally allow for a type-1 AGN component, while the yellow squares use \texttt{eazy} photometric redshifts and a type-2 AGN template only.} Overall, we find good agreement among the three luminosity functions, with the largest discrepancies appearing at the highest luminosities, where the measurements are most affected by small-number statistics.}

        \label{fig:app_LF}
    \end{minipage}

\end{figure*}

Figure~\ref{fig:app_LF} presents the non-parametric luminosity function and the corresponding number counts in each luminosity bin for $2.5 < z < 3.5$. The \tb{yellow diamonds} show the fiducial luminosity function adopted in this work: spectroscopic redshifts are fixed for the 19 sources where they are available, while photometric redshifts for the remaining objects are derived with \texttt{CIGALE}. The \tb{squares} follow the same treatment, again fixing the 19 spectroscopic redshifts, but computes the photometric redshifts with \texttt{eazy} for the rest. Although a few sources shift between luminosity bins—most noticeably at the low-luminosity end, the overall luminosity function remains robust to the choice of photo-$z$ method, with all measurements consistent within the $1\sigma$ uncertainties.

The \tb{circles} show the luminosity function obtained when fitting both type-1 and type-2 AGN, rather than the fiducial type-2 only approach used throughout the main paper. We again find excellent agreement, with results consistent within the error bars ($1 \sigma$). The modest differences arise from how the bolometric luminosity is partitioned: in the type-1 fits, more optical/UV emission is attributed to the AGN component, which changes the relative weight of the mid-IR constraints. However, as discussed before (\S~\ref{sec:CIGALE-setup}), the optical/UV constraints are strongly degenerate with the stellar component. Because this choice does not produce a statistically significant change, we adopted the type-2 only fitting in the main analysis.

\section{Comparison with the AGN identified in the JWST SMILES program} \label{sec:app_Lyu}

\begin{figure*}[!h]
    \includegraphics[width=\textwidth]{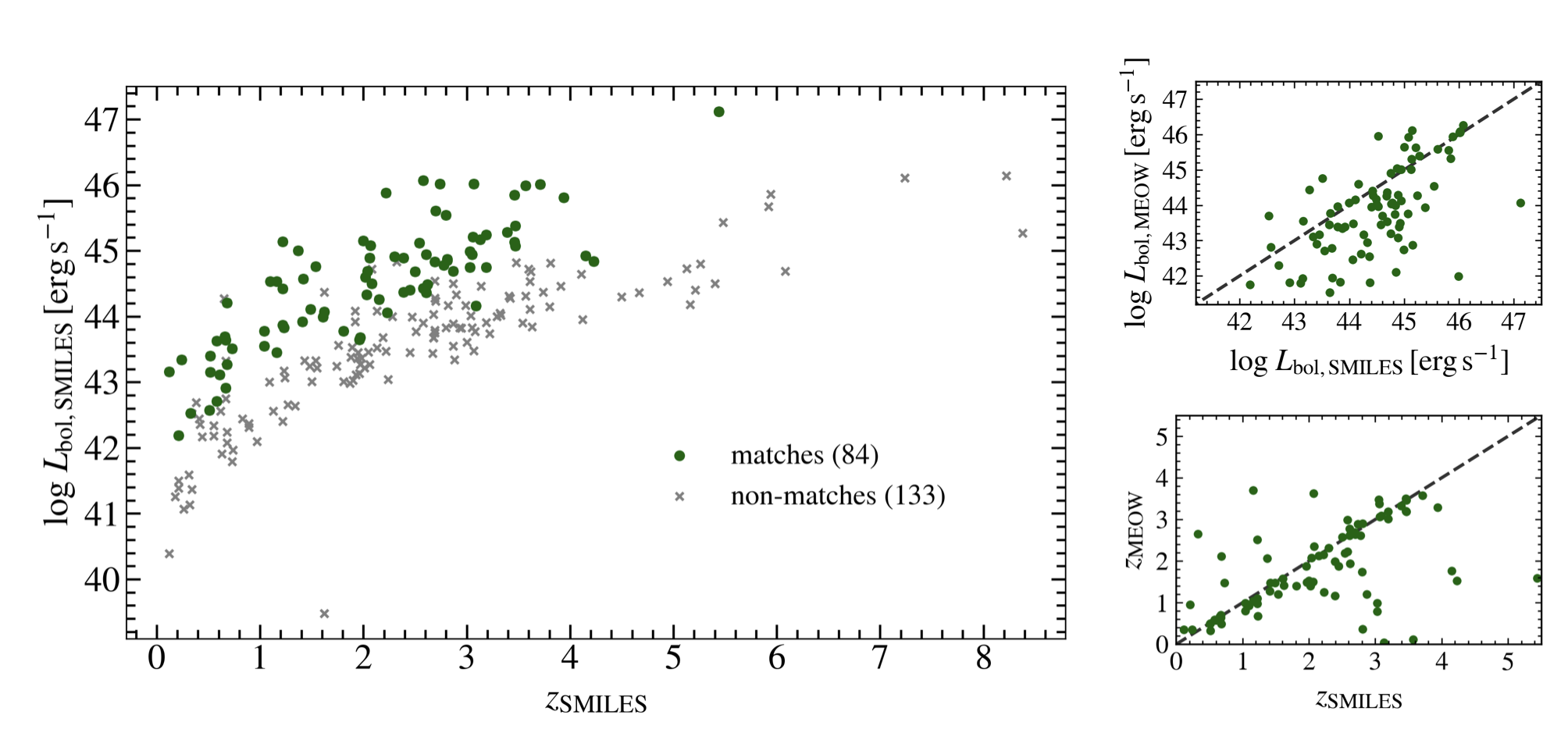}
    \caption{\textbf{Comparison between the \cite{Lyu2024} AGN sample and this work, in the overlapping footprint.} Out of the 217 AGN identified in SMILES \citep{Lyu2024}, we find 84 counterparts in our sample. The unmatched AGN are predominantly at the low luminosity end, with six exceptions: two lack F444W coverage and are therefore not identified in our sample, while four are affected by different deblending parameters in this work that result in only the brighter SMILES component being detected. Among the matched sources, we show the comparison between our luminosity and redshift estimates vs the SMILES counterparts, on the right. }
    \label{fig:comp_Lyu}

\end{figure*} 

Given the overlap of our analysis with the SMILES footprint, we cross-matched our AGN sample with the AGN reported in \cite{Lyu2024}. We note that there are some significant differences in the methodologies of the two studies however. First, we require a detection in all F444W, F1000W, and F2100W in this work for the sources, meaning for example that we do not include sources without F444W coverage, even if they show a clear MIRI detection. Second, the sample in \cite{Lyu2024} includes six additional MIRI filters (F560W, F770W, F1280W, F1500W, F1800W, F2550W), providing additional depth than F1000W and F2100W alone. This allows them to probe lower luminosities and identify more sources that may be detected in other MIRI bands but not in F1000W and/or F2100W.

Additionally, the SED fitting methodologies are different between the two studies.  In \cite{Lyu2024}, the SED fitting was performed with a modified Prospector framework that jointly models stellar emission, dust-obscured star formation, and AGN components, building on the Bayesian stellar-population inference framework of Prospector and FSPS (e.g., \citealt{Conroy2009,Leja2019,Johnson2021}). The fits were explored with dynamic nested sampling, and AGN classifications were retained only when the posterior solutions were well behaved and consistent with the source photometry and morphology.

We show the results of the cross-match in Fig.~\ref{fig:comp_Lyu}. We recover 84 counterparts out of the 217 SMILES AGN. As expected, the matched sources are those with bolometric luminosities above the F1000W and/or F2100W detection limits, while the majority of the unmatched objects fall below our sensitivity threshold. After visual inspection of all the cut-outs, we identify six exceptions in which the lack of a match is not driven by sensitivity: two sources fall outside the F444W coverage, and four are near a bright source and are not deblended by our source detection procedure. We note however that our source injection simulations account for such incompleteness.

In the right panel of Fig.~\ref{fig:comp_Lyu}, we compare the bolometric luminosities and redshift estimates derived by MEOW and SMILES. Our SED fitting generally yields lower bolometric luminosities, with a median of 0.5 dex difference than reported by SMILES. Importantly, most of the sources driving these discrepancies lie outside the luminosity range that enters our luminosity function and obscured-fraction analysis (\S~\ref{Sec:results}), limiting their impact on the main results of this study. Given that both measurements rely on SED-fitting methodologies, we do not attempt to determine which set of luminosities is more accurate; instead, we interpret the offsets as reflecting known systematic uncertainties in AGN SED modeling (e.g., \citealt{Buchner2024}). In this sense, our systematically lower bolometric luminosities would, if anything, render our inferred obscured fractions conservative. The redshift comparison shows substantially better overall agreement, with a small number of extreme outliers reaching $|\Delta z|\approx 4$. At the population level, these differences are not expected to materially affect our statistical conclusions (see also Appendix~\ref{sec:app_LF}).

Lastly, we note that 28 AGN in our sample used for the luminosity function analysis (\S~\ref{Sec:LF}) and lying within the SMILES footprint are not classified as AGN in SMILES, which we attribute primarily to differences in the adopted SED modeling. \tb{We note that these 28 sources are uniformly spread in redshift, but tend to be lower luminosities than the other MEOW AGN in the same redshift bins.} Furthermore, only 46 of the 84 cross-matched sources are independently identified as AGN in our analysis (i.e., $\fracA > 0.1$). These systematic differences support the conclusion that our luminosity function and obscured fraction measurements are robust to reasonable variations in SED assumptions, but highlight the large uncertainties in inferring individual quantities via SED fitting.
\bibliography{references}{}
\bibliographystyle{aasjournal}
\end{CJK*}
\end{document}